# Group theoretical analysis of structural instability, vacancy ordering and magnetic transitions in the system troilite (FeS) – pyrrhotite (Fe$_{1-x}$S)

Authors

**Charles Robert Sebastian Haines**[a]*, **Christopher J. Howard**[b], **Richard J. Harrison**[a] and **Michael A. Carpenter**[a]

[a]Department of Earth Sciences, University of Cambridge, Downing Street, Cambridge, Cambridgeshire, CB3EQ, UK

[b]School of Engineering, The University of Newcastle, University Drive, Callaghan, NSW, 2308, Australia

Correspondence email: crsh2@cam.ac.uk

**Funding information**   Leverhulme Trust.

**Synopsis**   A group-theoretical framework to describe vacancy ordering and magnetism in the Fe$_{1-x}$S system is developed.

**Abstract**   A group-theoretical framework to describe vacancy ordering and magnetism in the Fe$_{1-x}$S system is developed. This framework is used to determine the sequence of crystal structures consistent with the observed magnetic structures of troilite (FeS), and to determine the crystallographic nature of the low-temperature Besnus transition in Fe$_{0.875}$S. We conclude that the Besnus transition is a magnetically driven transition characterised by the rotation of the moments out of the ac-plane, accompanied by small atomic displacements that lower the symmetry to triclinic at low temperatures. Based on our phase diagram, we predict related magnetically driven phase transitions at low temperatures in all the commensurate superstructures of pyrrhotite. The exact nature of the transition is determined by the symmetry of the vacancy ordered state Based on this we predict spin-flop transitions in 3C and 5C pyrrhotite and a transition akin to the Besnus transition in 6C pyrrhotite. Furthermore, we clarify that 3C and 4C pyrrhotite carry a ferrimagnetic moment whereas 5C and 6C are antiferromagnetic.

# 1. Introduction

The mineral pyrrhotite has compositions that are usually expressed in terms of an ideal formula, $Fe_{1-x}S$ ($0 \leq x \leq 0.15$). Pyrrhotite adopts the hexagonal NiAs structure at high temperatures and has an increasing concentration of vacancies on the metal cation site with increasing x. Such a simple description belies the remarkable diversity of superstructures and phase transitions that are observed in both natural and synthetic samples, however. The phenomenological richness is reflected in a complex subsolidus phase diagram (Figure 1; Grønvold and Stolen 1992). The hexagonal structure at high temperatures is labelled as 1C. This gives way to four different commensurate superstructures with repeats along the crystallographic c-axis, corresponding to 3, 4, 5 and 6 times that of the parent structure (3C, 4C, 5C, 6C), and incommensurate superstructures in which a non-rational repeat varies with composition and temperature. Although other compilations differ in detail (Wang and Salveson 2005; Bennett and Graham 1980), it has long been understood that these complex structural relationships arise as a consequence of Fe/vacancy ordering, which, at particular stoichiometries, produce vacancy ordered phases that are more stable than the disordered continuous solid solution (Morimoto & Nakazawa, 1968; Morimoto et al., 1970; Carpenter & Desborough, 1964; Andresen et al., 1960; Andresen et al., 1967). In addition, FeS (troilite) has no vacancies but has at least four potential structural instabilities with symmetries relating to the K, M, H and L points of the Brillouin zone of the parent primitive hexagonal lattice (Ricci and Bousquet, 2016).

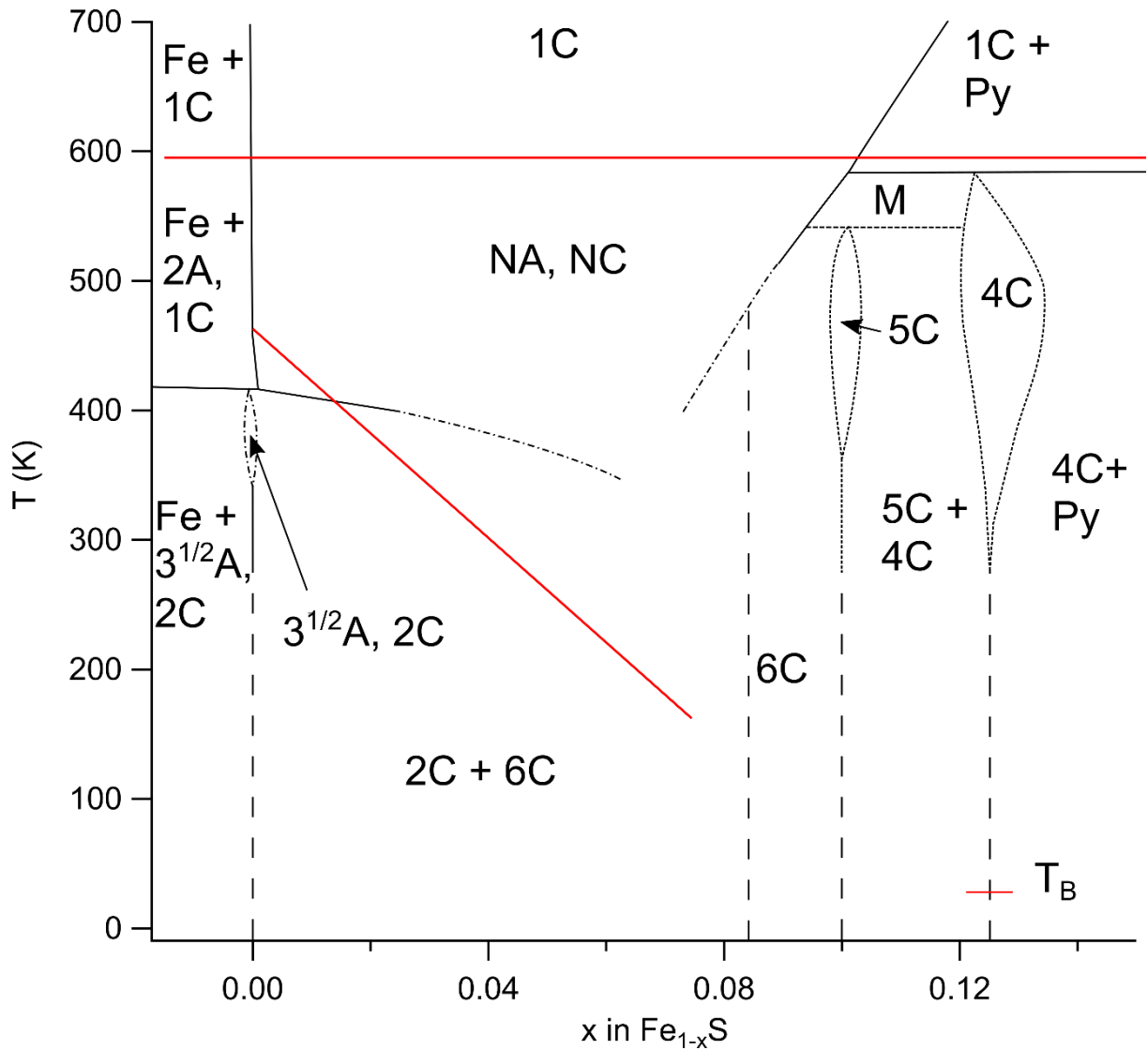

**Figure 1** Schematic phase relations, following Grønvold and Stolen (1992), for the system $Fe_{1-x}S$. A and C represent lattice parameters of the parent hexagonal NiAs structure (1C). Commensurate superstructures have multiple repeats of these. Incommensurate superstructures can occur at a wide range of compositions and temperatures, including in the field labelled M (Nakazawa and Morimoto 1971). The stability field at intermediate compositions and temperatures, labelled *N*A, *N*C by Grønvold and Stolen (1992), is shown as 1C by Wang and Salveson (2005). Selected magnetic transitions have been added. The red line running horizontally across the whole phase diagram is the magnetic ordering temperature, known in FeS as the β-transition. The red diagonal line shows the FeS spin-flop (Morin) transition at ~450 K, reducing to ~170 K at x = 0.07 (Horwood et al. 1976). $T_B$ and the short red line at ~35 K on the 4C line mark the Besnus transition. The line phases for FeS as well as 4C and 5C have been extended to low temperature and a line at the ideal stoichiometry ($Fe_{11}S_{12}$) of 6C has been added (- - -).

An equivalent diversity of magnetic structures has also been observed in natural and synthetic pyrrhotite (e.g., Andresen and Torbo 1967, Schwartz and Vaughan 1972, Bennett and Graham 1980,

Wang and Salveson 2005, Pearce et al. 2006). The first magnetic transition with falling temperature occurs at ~590 K and is from paramagnetic to antiferromagnetic in FeS and is from paramagnetic to ferrimagnetic or antiferromagnetic across the rest of the solid solution. Additional spin-reorientation transitions occur at lower temperatures and show a high sensitivity to structure type. The best characterised of the low-temperature transitions are a spin-flop (Morin type) transition at ~450 K in hexagonal FeS (Sparks et al 1962; Andresen and Torbo 1967; Takahashi 1973; Horwood et al 1976; Grønvold, and Stolen 1992), and the Besnus transition at ~35 K in monoclinic $Fe_7S_8$ (Besnus and Meyer 1964; Fillion and Rochette 1988; Dekkers et al 1989; Rochette et al. 1990). The Morin transition is a reorientation of individual moments from parallel to the crystallographic c-axis below the transition temperature to perpendicular to the c-axis above it. The transition temperature reduces with decreasing Fe content (Horwood et al. 1976). The Besnus transition involves more subtle changes of spin orientation in the 4C superstructure (Powell et al. 2004; Wolfers et al. 2011; Kind et al. 2013; Koulialias et al. 2016; Volk 2016; Koulialias et al. 2018) and is not observed in crystals with similar composition when the superstructure type is 3C (Horng and Roberts 2018). Some of these transitions have been added to Figure 1, but the magnetic properties of all the superstructure types have not yet been determined.

In a geological context, pyrrhotite with stoichiometry ~$Fe_7S_8$ is an important carrier of palaeomagnetic remanence in terrestrial and extraterrestrial rocks and has been proposed to be responsible for the strong remanent magnetic anomalies observed in the Martian crust (Rochette et al 2005; Louzada et al 2007). Its remarkable mix of structural transitions, vacancy ordering and magnetic properties, with the further possibility of a magnetoelectric effect (Ricci and Bousquet 2016), makes it of interest in the broader context of multiferroic materials (Eerenstein et al 2006) and in the topical field of domain boundary engineering (Salje 2009 and 2010). Phenomenological mean-field theories provide an effective theoretical framework for relating structural states and physical properties to external fields because of their rigour in defining the form and consequences of coupling between multiple order parameters. Underpinning this approach is the fundamental role of symmetry, as has been used to describe combinations of ferroelectric displacements, octahedral tilting, magnetic ordering, cation order and Jahn-Teller distortions in perovskites (Howard and Stokes 1998; Howard et al. 2003; Howard and Stokes 2004; Stokes et al. 2002; Carpenter and Howard 2009; Howard and Carpenter 2010; Senn et al. 2018).

An initial objective of the present study was to define the relationships between the crystallographic and magnetic structures of pyrrhotite from the perspective of symmetry. The parental structure is taken to be FeS with the NiAs structure, space group P6$_3$/*mmc* (Fig. 2a) and the Brillouin zone for a hexagonal P lattice (Fig. 2b). The Fe atom is set at Wyckoff 2*a*, 0, 0, 0 and the S atom at Wyckoff 2d, 1/3, 2/3, 3/4. Representative lattice parameters, determined for $Fe_{0.88}S$ at 773 K (Powell et al., 2004) are *a* = 3.5165 Å, *c* = 5.7142 Å. The structure comprises a distorted hexagonal close packed array of S atoms with the Fe atoms located in the octahedral sites.

The distortions to be considered in this work are of three types: i) purely displacive distortions; ii) vacancy ordering; and iii) magnetic ordering of moments on the Fe atoms. We attempt to associate an irreducible representation (irrep) and corresponding order parameter with each distortion encountered. To this end we make use of the group theoretical tools available in the ISOTROPY Software Suite (Stokes and Hatch 1988), making particular use of ISOCIF, ISOTROPY, ISODISTORT (Campbell *et al*., 2006) and INVARIANTS (Hatch & Stokes, 2003). Once the primary and secondary irreps, have been identified (we use the notation of Miller & Love, 1967, throughout) we list the space groups of the diverse distorted structures, and examine the couplings between the different distortions involved. We demonstrate that the known superstructure types can be understood in terms of irreps associated with the line between L and M points of the Brillouin zone (Fig. 2b), while magnetically ordered structures can be understood in terms of magnetic irreps associated with the Brillouin zone centre. Analysis of the primary and secondary irreps, and the manner in which they can combine, leads to a model for the Besnus transition in 4C pyrrhotite and to a prediction of the likely magnetic behaviour of crystals with different Fe/vacancy ordering schemes. A set of Landau expansions to account for coupling between order parameters, and with strains which accompany the various phase transitions, is included in Appendix B. The magnetoelastic properties of single crystals with 4C and 5C superstructures are set out in two separate experimental papers (Haines et al …..).

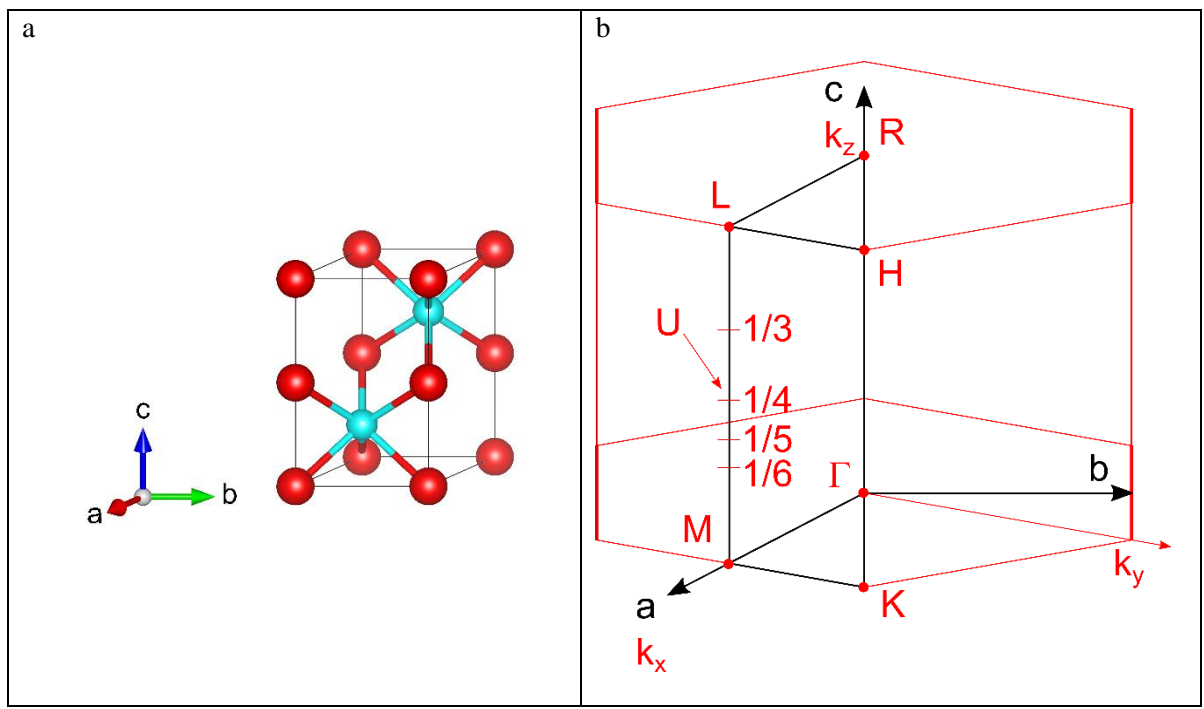

**Figure 2** (a) NiAs parent structure of pyrrhotite, stable above $T_\beta$~ 590 K. Hexagonally close packed sulphur atoms are shown in blue. Fe atoms filling the octahedral sites are shown in red. (b) Brillouin zone for the hexagonal P lattice, with labelling of high symmetry points (filled red circles) and the U-line used in the group theoretical analysis. The directions corresponding to the real space unit cell are shown in black. Points on the U-line used in this work ($k_z$ = 1/3, 1/4, 1/5 and 1/6) are also shown.

## 2. Structural relationships and vacancy ordering

We start our analysis with the best known of the pyrrhotite structure: 4C. After describing the vacancy ordering in this system in terms of group theory, we apply the same methods to the other pyrrhotite superstructures (working from high to low vacancy concentration) before discussing the case of troilite, FeS, in which there are no vacancies.

### 2.1. 4C structure

The room temperature structure of 4C pyrrhotite has crystallographic space group *C*2/*c* (Fig. 3; Powell et al. 2004). Close packed layers of S in the ab-plane alternate with layers of Fe atoms. Vacancies are located in alternate layers (of Fe atoms), and within these layers they are found on alternate sites in alternate (close-packed) rows (Powell *et al.*). This means that, within the layers containing vacancies, one site in four is vacant. There are four distinct sites for the vacancies – when referred to the parent structure a vacancy can be located above the 0,0; 1,0; 1,1 or 0,1 positions on the base of the hexagonal unit cell. The layers containing vacancies are stacked up the hexagonal axis in a sequence involving all four distinct vacancy sites (e.g. ABCD), leading to a structure that has a repeat along that axis four times that of the hexagonal parent. This description of the vacancy ordering scheme in terms of stacking up vacancy containing and non-vacancy containing layers was first used by Bertaut (1953) for the 4C structure and subsequently developed for the 3C (Fleet 1971, Nakano et. al. 1979) and 6C (Koto et. al. 1975) structures before being discussed in detail more generally by Yamamoto and Nakazawa (1982). Bertaut (1953) first described the structure (Fig. 3) in an unconventional space group *F*2/*d*, on a nearly orthorhombic $2\sqrt{3}a_p$ by $2a_p$ by $4c_p$ cell, and it was subsequently refined from X-ray data in the same unconventional setting by Tokonami *et al.* (1972). It was recognised that the 4C structure could be described in the conventional space group *C*2/*c* on a monoclinic cell approximately $2\sqrt{3}a_p$ by $2a_p$ by $\sqrt{3a_p{}^2 + 4c_p{}^2}$ with $\tan(180 - \beta) = 2c/\sqrt{3}a$ (β≈118º), and the structure was refined in this setting by Powell *et al.* (2004).

Every second Fe layer contains vacancies in an ordered arrangement such that the repeat distance in the c* direction is four times that of the parent NiAs structure. The relationship to the parent structure is most easily visualised if the alternative setting of space group *F*2/*d* is used, see Fig. 3, in which case the lattice parameters are ~2*A*, 2√3*A*, 4*C*, ▯ ~90°.

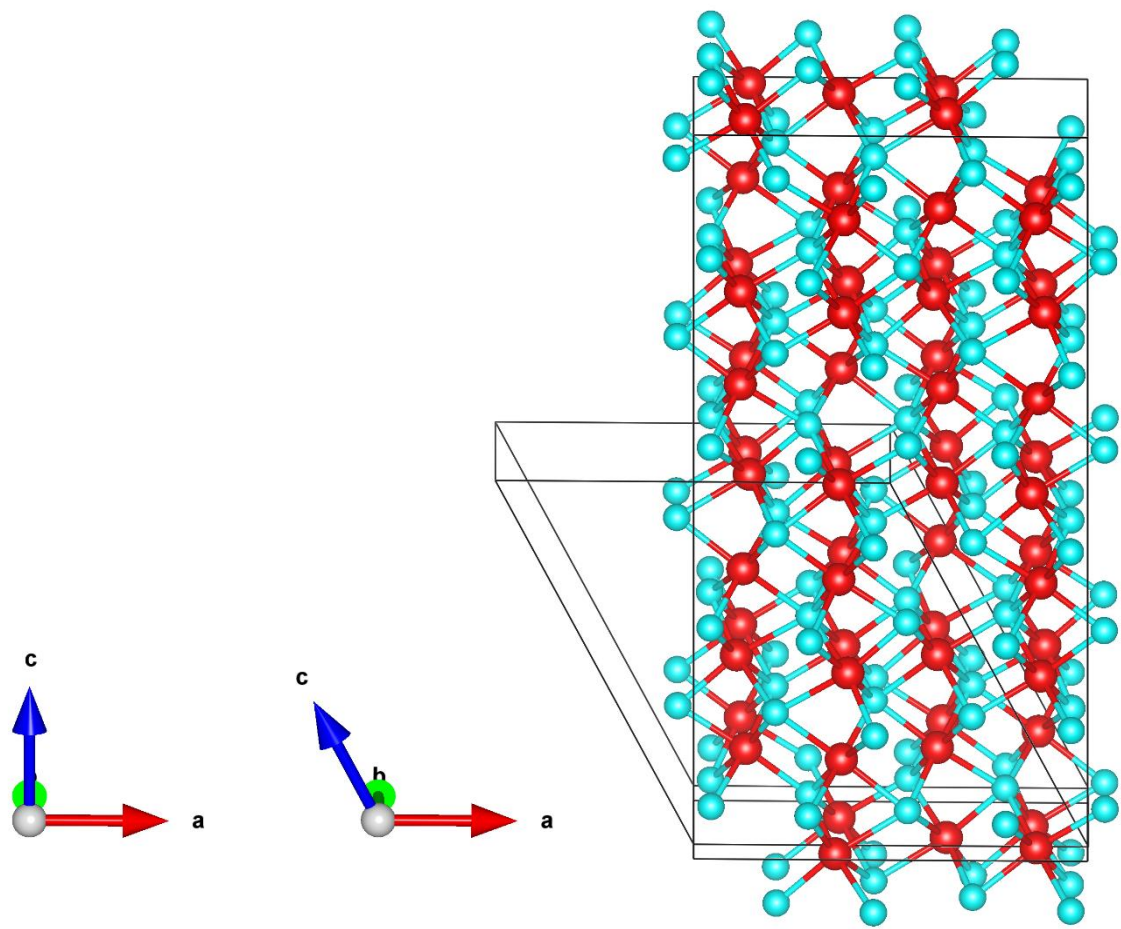

**Figure 3** A single unit cell of the structure of 4C $Fe_7S_8$ at room temperature, after Powell et al (2004). Sulphur atoms are shown in blue. Fully occupied Fe layers alternate with layers in which the vacancies are ordered. The unit cell of the standard setting is also shown. There is a translation of (0,1/2,0) between the two cells. The left axes relate to the unconventional F2/d cell and the right axes to the conventional C2/c cell.

The order parameter which gives rise to the symmetry change, $P6_3/mmc \rightarrow C2/c$, has the symmetry properties of U1(1/2,0,1/4). It may be helpful to note that a distortion with this k-vector (1/2,0,1/4) can be thought of as having two distinct effects: the first component ($k_x$=1/2) gives the vacancy order within the plane and the third component ($k_z$=1/4) leads to the superstructure repeat along the c-axis of the parent. Table 1 lists all possible subgroups, from which it is clear that the order parameter direction is P4, (*a*,0,0,*a*,0,0). The basis vector is (2,-2,0), (2,2,0), (-1,1,2) and the origin is at (0,1/2,0). U4 is a secondary order parameter in this case, and the same result would be obtained if U4 was taken to be primary and U1 to be secondary. U1 and U4 do not appear to be interchangeable in this way for all values of *g*, however, and it is therefore assumed that U1 provides the fundamental primary order parameter.

It is now necessary to show that U1(P4) does indeed lead to the observed/reported ordering scheme. U1(P4) can impact on the site occupancies, but it does not by itself lead to the observed disposition of vacancies. However, the space group symmetry allows a number of secondary distortions, some of which also affect the occupancies. It is necessary to invoke these occupation-affecting secondary modes, with amplitudes tied to the amplitude of U1(P4), to match the observed disposition of vacancies. In another context these secondary modes have been described as 'dependent modes' (Campbell *et al.*, 2018). An example of the procedure used to determine the dependent modes using ISOTROPY is given in Appendix A. Firstly, all secondary modes related to the symmetry change *P*6$_3$/*mmc* → *C*2/*c* (P4) were examined to determine which ones define atomic occupancy (Appendix A). Each of these secondary irreps was then examined, as illustrated in Figure A2, to find the set of ratios for their individual contributions which would yield the observed vacancy distribution. Contributions from irreps U4, L1, $G_3^+$ and $G_1^+$ in the proportions $\frac{U1-U4}{4\sqrt{2}} + \frac{L1}{8} + \frac{\Gamma_3^+}{8} + \frac{7\Gamma_1^+}{8}$ give the pattern of vacancy ordering in the 4C structure. The degree of order scales with the magnitude of the primary order parameter.

**Table 1** Order parameter components and unit cell configurations for subgroups of *P*6$_3$/*mmc* which can arise from phase transitions in which irrep U1(1/2,0,1/4) is the active representation. The table has been curtailed at the final single component order parameter distortion, P8. See Table S1 for the full table.

| Space Group | Space Group Number | OPD Name | OPD vector | Basis Vectors | Origin |
|---|---|---|---|---|---|
| C2/m | 12 | P1 | (a,0,0,0,0,0) | (2,1,-4),(0,1,0),(2,1,0) | (0,0,0) |
| Imm2 | 44 | P2 | (a,-0.414a,0,0,0,0) | (0,0,4),(0,1,0),(-2,-1,0) | (-7/8,1/16,-7/4) |
| C2/m | 12 | P3 | (a,0,a,0,0,0) | (2,-2,0),(2,2,0),(-1,1,2) | (0,0,0) |
| C2/c | 15 | P4 | (a,0,0,a,0,0) | (2,-2,0),(2,2,0),(-1,1,2) | (0,1/2,0) |
| Fmm2 | 42 | P5 | (a,-0.414a,a,-0.414a,0,0) | (0,0,4),(2,2,0),(-2,2,0) | (-1/8,1/8,1/4) |
| Fdd2 | 43 | P6 | (a,-0.414a,0.414a,a,0,0) | (0,0,4),(2,2,0),(-2,2,0) | (-7/8,7/8,7/4) |
| P$\bar{3}$m1 | 164 | P7 | (a,0,a,0,a,0) | (2,0,0),(0,2,0),(0,0,4) | (0,0,0) |
| P$\bar{6}$m2 | 187 | P8 | (a,-0.414a,a,-0.414a,a,-0.414a) | (0,-2,0),(2,2,0),(0,0,4) | (0,0,1/4) |

## 2.2. 3C structure

3C pyrrhotite is metastable at room temperature and is obtained by quenching crystals with compositions in the vicinity of ~$Fe_7S_8$ from above ~600 K (Fleet 2006). Nakano et. al. (1979) were able to refine the structure for a $Fe_7S_8$ crystal that had been quenched from 300 °C as a single phase. They found it to be on a hexagonal cell of dimensions approximately 2A by 3C and concluded that the space group was *P*$3_1$21 (#154), as did Keller-Besrest et al. (1982). As in the 4C structure, vacancies occur on alternate layers, and within the vacancy-containing layers the arrangements are the same as in the 4C structure. The difference is in the stacking of these layers. Again, there are four distinct possible sites for the vacancies, but in this case the layers are stacked so only three of these are vacant. When referred to the parent structure, the vacancies are located successively above the 1,0; 0,1 and -1,-1 (equivalent to 1,1) positions on the base of the hexagonal unit cell; there are no vacancies above the 0,0 position. We concur that this structure is better described in *P*$3_1$21 than in *P*$3_1$ used by other authors (Fleet, 1971); all positions in *P*$3_1$ are general and Fleet suggests an arrangement of vacancies on a subset of these positions, whereas in *P*$3_1$21 we can achieve the postulated structure by declaring vacancies to be on one set of Wyckoff 3*a* sites.

The ***k***-vector that will describe this arrangement of vacancies is ***k*** = 1/2,0,1/3, again on the U line of symmetry. It follows that U1 and U4 are again the irreps of interest. Starting with irrep U1, we can list details of the structures arising for different 'simple' order parameter directions (OPDs), Table 2. We find one structure in *P*$3_1$21, from U1(P7), on a basis (2,0,0), (0,2,0), (0,0,3) and at origin (0,1/2,0). This makes the cell dimensions 2A by 2A by 3C, and we have the structure described just above. It may be worth pointing out that using the P4 direction that describes the distortion in the 4C case gives a space group of P$\bar{6}$m2.

**Table 2**  Order parameter components and unit cell configurations for subgroups of P$6_3$/mmc which can arise from phase transitions in which irrep U1(1/2,0,1/3) is the active representation. The table has been curtailed at the final single component order parameter distortion, P8. See Table S10 for the full table.

| Space Group | Space group Number | OPD Name | OPD vector | Basis Vectors | Origin |
|---|---|---|---|---|---|
| 58 | Pnnm | P1 | (a,0,0,0,0) | (0,0,3),(2,1,0),(0,1,0) | (0,0,0) |
| 59 | Pmmn | P2 | (0,a,0,0,0) | (0,0,3),(0,-1,0),(2,1,0) | (1/2,1/2,0) |
| 164 | P$\bar{3}$m1 | P3 | (a,0,a,0,a,0) | (2,0,0),(0,2,0),(0,0,3) | (0,0,0) |
| 187 | P$\bar{6}$m2 | P4 | (a,-0.577a,a,-0.577a,a,-.577a) | (0,-2,0),(2,2,0),(0,0,3) | (0,0,1/4) |
| 64 | Cmca | P5 | (a,0,0,0,a,0) | (2,0,0),(2,4,0),(0,0,3) | (0,0,0) |
| 63 | Cmcm | P6 | (0,a,0,0,0,-a) | (2,0,0),(2,4,0),(0,0,3) | (1/2,0,0) |
| 152 | P3$_1$21 | P7 | (a,1.732a,-2a,0,a,-1.732a) | (2,0,0),(0,2,0),(0,0,3) | (0,0,1/2) |
| 151 | P3$_1$12 | P8 | (a,0.577a,a,0.577a,0,-1.155a) | (0,-2,0),(2,2,0),(0,0,3) | (0,0,3/4) |

To check that we can reproduce the vacancy ordering scheme, we follow exactly the same procedure as in the 4C case. We find that the combination of model amplitudes indicated by $\frac{(\sqrt{3}U4-U1)}{12} + \frac{M1-M3}{24} - \frac{3\Gamma_3^+}{12} + \Gamma_1^+$ gives the vacancy ordering arrangement as described.

**2.3. 5C structure**

De Villiers (2009) reports a Cmce (Cmca) structure (based on the earlier model of Morimoto et. al. 1975), Elliot (2010) a P2$_1$/c structure, and finally, Liles and De Villiers (2012) a P2$_1$ structure. All of these structures are in the table of possible subgroups arising from a U line distortion (Table S4). The most recent determination of the structure of 5C pyrrhotite is in space group $P2_1$ (Liles and de Villiers 2012) using a setting with cell parameters corresponding to 2A, 5C, 2A, □ ~ 120°. We have not determined the necessary combination of irreps that give the vacancy distribution reported. There remains significant uncertainty over the vacancy distribution. In all the other systems (3C, 4C and 6C) the occupancies are set at either 1,0 or ½ and can then be exactly described by occupancies on specific Wyckoff sites. If there is a distribution of occupancy akin to an 'occupancy wave', then a description starting from an incommensurate vacancy ordering structure may be more illuminating. Table 3 gives distortions described by a single component order parameter.

**Table 3** Order parameter components and unit cell configurations for subgroups of *P*6$_3$/*mmc* which can arise from phase transitions in which irrep U1(1/2,0,1/5) is the active representation. The table has been curtailed at the final single component order parameter distortion, P6. See Table S4 for the full table.

| Space Group | Space group Number | OPD Name | OPD vector | Basis Vectors | Origin |
|---|---|---|---|---|---|
| 58 | Pnnm | P1 | (a,0,0,0,0,0) | (0,0,5),(2,1,0),(0,1,0) | (0,0,0) |
| 59 | Pmmn | P2 | (0,a,0,0,0,0) | (0,0,5),(0,-1,0),(2,1,0) | (1/2,1/2,0) |
| 164 | P$\bar{3}$m1 | P3 | (a,0,a,0,a,0) | (2,0,0),(0,2,0),(0,0,5) | (0,0,0) |

| | | | (a,-0.325a,a,-0.325a,a,- | | |
|---|---|---|---|---|---|
| 187 | P$\bar{6}$m2 | P4 | 0.325a) | (0,-2,0),(2,2,0),(0,0,5) | (0,0,1/4) |
| 64 | Cmca | P5 | (a,0,0,0,a,0) | (2,0,0),(2,4,0),(0,0,5) | (0,0,0) |
| 63 | Cmcm | P6 | (0,a,0,0,0,-a) | (2,0,0),(2,4,0),(0,0,5) | (1/2,0,0) |

**2.4. 6C structure**

The 6C structure was first confirmed by single crystal XRD by Koto et. al. (1975) on a crystal of stoichiometry $Fe_{11}S_{12}$. They proposed a basis of (2,0,0) (2,4,0) (0,0,6) and two possible unconventional space groups of Fd (conventional equivalent is Cc) or F2/d (C2/c). Koto et al. (1975) adopted the former, because it gave 12 independent Fe positions and the possibility of keeping just one of these vacant. However, their analysis suggested that two of these twelve are half occupied, and therefore that the structure is F2/d. Choosing Fd forces a subset of a Wyckoff set to have different occupancies from the rest of that set, which is not justifiable. On the basis of this symmetry, the U1 irrep in direction P4 (Table 4) gives the correct distortion and the ordering scheme of the vacancies (or in this case half-filled sites) matches the observed. The Cc structure is obtained from the U1 irrep with one more independent parameter in 2 ways: C7 and C9 (see Table S7). The combination of irreps needed to give the reported distribution of sites on the C2/c setting with partial Fe/vacancy order is

$\frac{(1+\sqrt{3})U1}{24} + \frac{(1-\sqrt{3})U4}{24} + \frac{\sqrt{3}U1(1/2,0,1/3)}{12} + \frac{\Delta 4}{4} - \frac{L1}{24} + \frac{11\Gamma_1^+}{12}$.

**Table 4** Order parameter components and unit cell configurations for subgroups of P63/mmc which can arise from phase transitions in which irrep U1(1/2,0,1/6) is the active representation. The table has been curtailed at the final single component order parameter distortion, P10. See Table S7 for the full table.

| Space group number | Space group | OPD name | OPD vector | Basis functions | Origin |
|---|---|---|---|---|---|
| 12 | C2/m | P1 | (a,0,0,0,0,0) | (2,1,-6),(0,1,0),(2,1,0) | (0,0,0) |
| 44 | Imm2 | P2 | (a,-0.268a,0,0,0,0) | (0,0,6),(0,1,0),(-2,-1,0) | (-11/12,1/24,-11/4) |
| 12 | C2/m | P3 | (a,0,a,0,0,0) | (2,-2,0),(2,2,0),(-1,1,3) | (0,0,0) |
| 15 | C2/c | P4 | (a,0,0,a,0,0) | (2,2,0),(-2,2,0),(1,1,3) | (0,1/2,0) |
| 43 | Fdd2 | P5 | (a,-0.268a,0.268a,a,0,0) | (0,0,6),(-2,2,0),(-2,-2,0) | (-4/3,-1/3,1) |
| 42 | Fmm2 | P6 | (a,-0.268a,a,-0.268a,0,0) | (0,0,6),(2,2,0),(-2,2,0) | (-1/12,1/12,1/4) |

| | | | | | | |
|---|---|---|---|---|---|---|
| 164 | P$\bar{3}$m1 | P7 | (a,0,a,0,a,0) | (2,0,0),(0,2,0),(0,0,6) | (0,0,0) |
| 187 | P$\bar{6}$m2 | P8 | (a,-0.268a,a,-0.268a,a,-0.268a) | (0,-2,0),(2,2,0),(0,0,6) | (0,0,1/4) |
| 152 | P3$_1$21 | P9 | (a,1.732a,-2a,0,a,-1.732a) | (2,0,0),(0,2,0),(0,0,6) | (0,0,1) |
| 151 | P3$_1$12 | P10 | (a,a,-1.366a,0.366a,0.366a, 1.366a) | (0,-2,0),(2,2,0),(0,0,6) | (0,0,5/4) |

## 2.5. Incommensurate structures

Incommensurate repeat distances in the range ~3.2C - 5.75C have been observed in crystals with a range of compositions at room temperature and in-situ at high temperatures. There is a tendency for the repeat distance to increase with decreasing vacancy concentration (Nakazawa and Morimoto 1971; Morimoto et al. 1975; Yamamoto and Nakazawa 1982). Yamamoto and Nakazawa (1982) proposed an incommensurate model with an 11C repeat for the structure of a crystal with composition $Fe_{0.91}S$ which had a repeat distance of 5.54. Use of a U1 irrep with $\mathbf{k} = (1/2,0,g)$ and an irrational value of $1/g$ gives the space group of the basic cell as *Cmcm*, in agreement with Yamamoto and Nakazawa (1982). Nakazawa and Morimoto (1971) also reported an incommensurate structure with a range of repeat distances in the ab-plane corresponding to ~40-90Å. This requires irreps associated with points close to the Γ-point along the line between Γ and M and is not considered further here.

## 2.6. Troilite, FeS

The paramagnetic–antiferromagnetic transition at $T_N$ ~590 K in troilite is accompanied by a change in crystallographic symmetry. An orthorhombic space group, *Pnma* ($a \sim C$, $b \sim A$, $c \sim \sqrt{3}A$), was proposed by King and Prewitt (1982) on the basis of refinement of diffraction data collected at 463 K, and a hexagonal space group, *P6$_3$mc* (2*A*, 1*C*), by Keller-Besrest and Collin (1990) on the basis of data collected at 453 and 429 K. The active irrep for both symmetry changes is $M_2^-$ and a list of possible subgroups for different order parameter directions is given in Table 5. The magnetic structure has individual magnetic moments aligned perpendicular to the c-axis from $T_N$ down to $T_S$ ~ 450 K, where a spin-flop (Morin) transition occurs and the orientation switches to being parallel to the c-axis (Hirahara and Murakami 1958; Andresen 1960; Sparks et al 1962; Andresen and Torbo 1967; Horwood et al 1976). An ordered arrangement of spins within the (001) plane would preclude hexagonal symmetry, favouring the orthorhombic structure, while spins aligned parallel to [001] would fit with hexagonal symmetry. Any distortions from hexagonal lattice geometry are small and the two reported structures may be correct for the temperatures at which the data were collected.

**Table 5** Subgroups of space group $P6_3/mmc$ obtained with $M_2^-$ as the active representation.

| OPD Label | OPD | Space Group Number | Space group | Basis Vectors | Origin |
|---|---|---|---|---|---|
| P1 | (a,0,0) | 62 | Pnma | (0,0,1),(0,-1,0),(2,1,0) | (1/2,0,0) |
| P2 | (a,0,a) | 64 | Cmca | (2,0,0),(2,4,0),(0,0,1) | (1/2,1,0) |
| P3 | (a,a,a) | 186 | P6$_3$mc | (2,0,0),(0,2,0),(0,0,1) | (0,0,0) |
| C1 | (a,0,b) | 14 | P2$_1$/c | (0,2,0),(0,0,1),(2,0,0) | (1/2,0,0) |
| C2 | (a,b,a) | 36 | Cmc21 | (2,0,0),(2,4,0),(0,0,1) | (0,0,0) |
| S1 | (a,b,c) | 4 | P2$_1$ | (0,2,0),(0,0,1),(2,0,0) | (0,0,0) |

The additional structural transition at ~420 K in troilite involves the symmetry change $P6_3mc$ – $P\bar{6}2c$ ($\sqrt{3}A$, $2C$) (Keller-Besrest and Collin 1990). $P\bar{6}2c$ is a subgroup of $P6_3/mmc$ but not of $P6_3mc$, so the transition is necessarily first order in character. The order parameter for $P6_3/mmc$ – $P\bar{6}2c$ belongs to the H point of the Brillouin zone, representing a different distortion to the U-line distortion being predominantly discussed in this article (Li and Frantzen 1994; Ricci and Bousquet 2016). There are three H-point irreps which will give the observed space group, although only H1 produces the reported atomic displacements, i.e. sulphur displaced in the *c*-axis and iron displaced within the *ab*-plane. Table 6 lists all structures arising from a distortion to the parent NiAs structure with the active irrep H1. The first entry is the structure reported for room temperature FeS.

**Table 6** Structures derived from NiAs structure of FeS with active irrep H1

| Space Group Number | Space group | OPD Label | OPD Vector | Basis Vectors | Origin |
|---|---|---|---|---|---|
| 190 | P$\bar{6}$2c | P1 | (0,0,0,a) | (2,1,0),(-1,1,0),(0,0,2) | (2/3,1/3,1/4) |
| 189 | P$\bar{6}$2m | P2 | (0,a,0,0) | (2,1,0),(-1,1,0),(0,0,2) | (2/3,1/3,3/4) |
| 15 | C2/c | P4 | (0,0,a,-a) | (1,2,0),(3,0,0),(0,0,-2) | (2,0,0) |
| 12 | C2/m | P8 | (a,a,0,0) | (1,2,0),(3,0,0),(0,0,-2) | (2,0,0) |
| 174 | P$\bar{6}$ | C1 | (0,a,0,b) | (2,1,0),(-1,1,0),(0,0,2) | (-1/3,1/3,3/4) |
| 9 | Cc | C2 | (0,0,a,b) | (1,2,0),(3,0,0),(0,0,-2) | (1,1,0) |
| 5 | C2 | C3 | (a,0,0,b) | (3,0,0),(1,2,0),(0,0,2) | (1/2,0,1/4) |

| | | | | | | |
|---|---|---|---|---|---|---|
| 8 | Cm | C4 | (a,b,0,0) | (1,2,0),(3,0,0),(0,0,-2) | (1,1,0) |
| 2 | P$\bar{1}$ | C10 | (a,a,b,-b) | (1,-1,0),(1,2,0),(0,0,2) | (1,1,0) |
| 5 | C2 | C11 | (a,-a,b,-b) | (1,2,0),(3,0,0),(0,0,-2) | (0,0,1/2) |
| 1 | P1 | 4D1 | (a,b,c,d) | (1,-1,0),(1,2,0),(0,0,2) | (0,0,0) |

There had been some discussion in the literature that FeS could be ferroelectric, implying that the room temperature structure has a polar space group (e.g., Van Den Berg et al. 1969; Li and Frantzen 1996), but Gosselin et al (1976) demonstrated that this is not the case.

Each of the main pyrrhotite superstructure types has lattice geometry that is only slightly distorted from that of the parent hexagonal structure. Their lattice parameters can be represented as being hexagonal, $2A$, $NC$, orthorhombic, $2A$, $2\sqrt{3}A$, $NC$ or monoclinic, $2A$, $2\sqrt{3}A$, $NC$, $\square \sim 90.0°$. From the perspective of group theory, the order parameter must have the symmetry properties of an irreducible representation (irrep) on the U line, $k = (1/2, 0, g)$, between the M point, $k = (1/2,0,0)$, and L point, $k = (1/2,0,1/2)$, of the Brillouin zone (Fig. 2b). Irrep U1 gives the correct structure in each case, with $g = 1/4$ for the 4C superstructure, 1/3 for the 3C superstructure, etc. This automatically generates secondary irreps that represent different symmetry adapted combinations of atomic displacements and/or atomic order. It has been found that the observed configurations of vacancies among the crystallographic sites depend on a combination of these secondary irreps in particular proportions. By way of contrast, and as previously shown by Li and Frantzen (1994), the $\sqrt{3}A$, 2C structure of FeS develops by the action of an order parameter which has the symmetry properties of an irrep at the H point (1/3,1/3,1/2) of the Brillouin zone (Ricci and Bousquet et al 2016).

**3. Magnetism**

Powell et al. (2004) determined that the magnetic unit cell of 4C pyrrhotite is the same as the crystallographic cell. Layers of Fe atoms in the ab-plane have moments ordered ferromagnetically with orientations which are reversed between alternate layers to give the antiferromagnetic structure. Ferrimagnetism occurs when vacancies order preferentially onto only one of the sets of layers. Our group theoretical analysis demonstrates that the magnetic ordering is associated with a combination of two Γ-point magnetic irreps, $m\Gamma_4^+$ and $m\Gamma_5^+$, of the space group of the parent structure, $P6_3/mmc$. In order to explore the implications of magnetic order parameters with these symmetries more widely, Table 7 includes magnetic structures that would arise as a consequence of magnetic order parameters with all allowed magnetic Γ-point irreps, $mG_2^+$, $mG_4^+$, $mG_5^+$ and $mG_6^+$. $mG_2^+$ allows alignment of moments parallel to the crystallographic c-axis, but only gives ferromagnetic structures, and $mG_6^+$ gives ferromagnetic structures with moments predominantly in the ab-plane. On their own, $mG_4^+$ and $mG_5^+$ would give antiferromagnetic structures that are hexagonal ($mG_4^+$), with moments perpendicular to the ab-plane, or orthorhombic ($mG_5^+$), with moments aligned within the ab-plane. The latter can also give a monoclinic structure in which a ferromagnetic component perpendicular to the ab-plane is allowed due to $mG_2^+$ as a secondary irrep. The magnetic gamma point distortions are shown in Fig. 4. The six subfigures correspond to the six independent order parameter directions allowed. The distortions characterised by a C1 order parameter direction in Table 7 are combinations of the two single parameter distortions allowed for that irrep.

**Table 7** Magnetic structures arising from Γ-point irreps of the parent *P*6$_3$/*mmc* space group. "Layers" refer to planes of Fe atoms perpendicular to the crystallographic c-axis. Crystallographic directions are given with respect to hexagonal cell.

| Active Irrep | OPD | Subgroup | Magnetic structure |
|---|---|---|---|
| $mG_2^+$ | P1 (a) | P6$_3$/mm'c' | FM moments parallel to c-axis |
| $mG_4^+$ | P1 (a) | P6$_3$/m'm'c | FM layers, moments parallel to c-axis, AFM between layers |
| $mG_5^+$ | P1 (a,0) | Cmcm | FM layers, moments parallel to [100], AFM between layers |
| | P2 (0,a) | Cm'c'm | FM layers, moments parallel to [120], AFM between layers. $mG_2^+$ as secondary order parameter gives FM component along c-axis |
| | C1 (a,b) | P2$_1$/m | FM layers, moments have components parallel to [100] and [120], AFM between layers. $mG_2^+$ as secondary order parameter gives FM component along c-axis |
| $mG_6^+$ | P1 (a,0) | Cm'cm' | FM, moments parallel to [120] |
| | P2 (0,a) | Cmc'm' | FM, moments parallel to [100] |
| | C1 (a,b) | P2$_1$/m' | FM, moments have components parallel to [100] and [120]. $mG_4^+$ as secondary order parameter gives AFM component along c-axis |

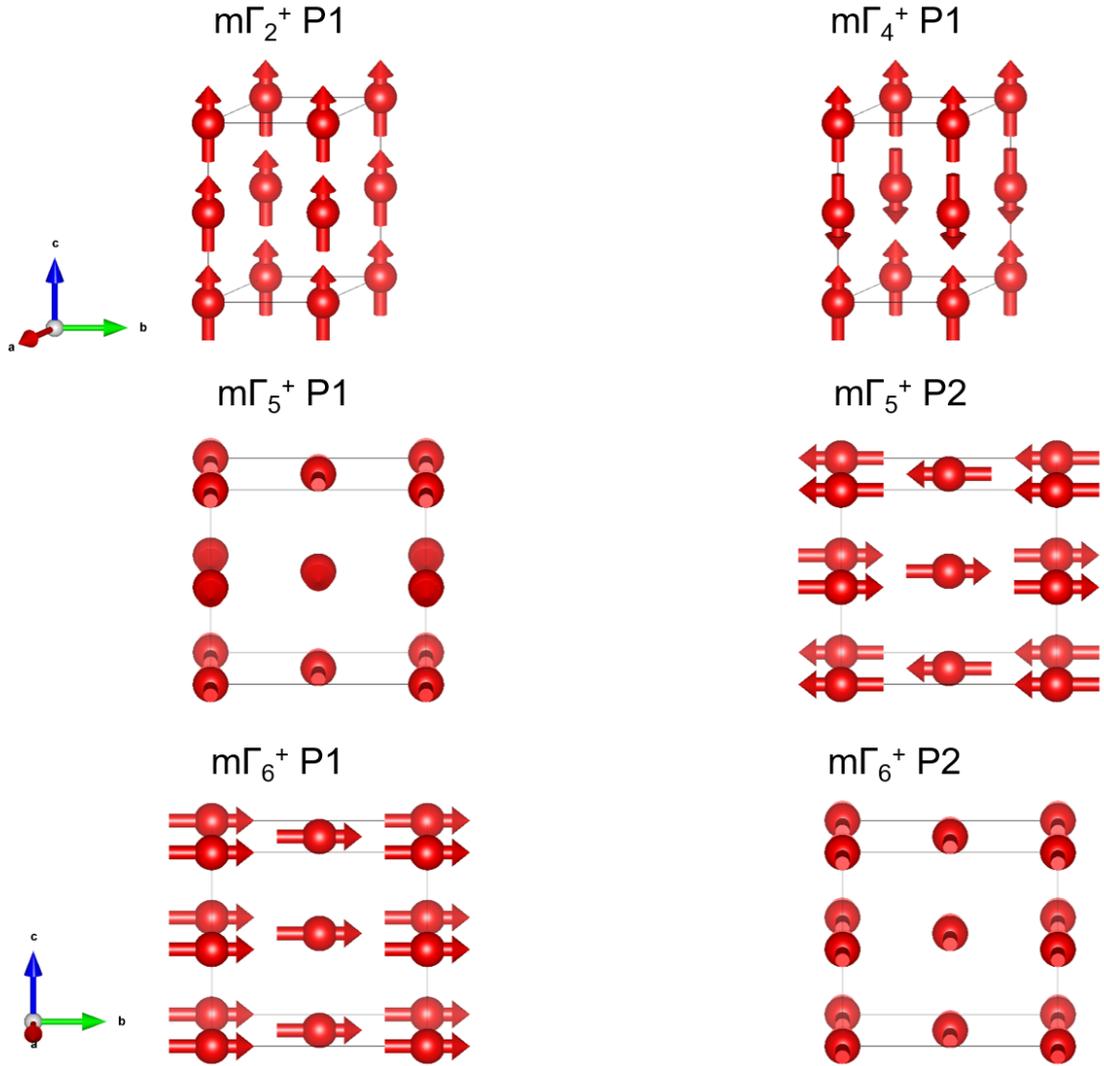

**Figure 4** Visualisation of all the allowed gamma point magnetic distortions. The sulphur atoms have been omitted for clarity. m$\Gamma_2^+$ and m$\Gamma_4^+$ do not change the unit cell from the parent NiAs structure (shown in outline). m$\Gamma_5^+$ and m$\Gamma_6^+$ lead to a cell doubling in the plane. All views are down the a*-axis with a 15° rotation around the b-axis.

### 3.1. FeS, troilite

The orientation of individual moments in troilite below $T_N$ is reported as being within the ab-plane (Hirahara and Murakami 1958; Andresen 1960; Sparks et al 1962; Andresen and Torbo 1967; Horwood et al 1976), requiring that the relevant magnetic irrep is $m\Gamma_5^+$. As set out in Table 8, combining $m\Gamma_5^+$ with $M_2^-$ gives two possibilities that are consistent with the *Pnma* structure reported by King & Prewitt (1982). These are *Pnma*, and *Pnm'a'* from directions P1 ($a$,-1.732$a$) and P2 ($a$,0.577$a$) of $m\Gamma_5^+$, respectively. Moments are lined up parallel to the crystallographic b-axis of the *Pnma* structure ([0$\bar{1}$0] of the parent hexagonal structure) or to the crystallographic c-axis of the *Pnm'a'* structure ([210] of the hexagonal parent). If the $mG_5^+$ direction is C1 (*a*,*b*), the magnetic space group is *P*2$_1$/*c* and the moments are not constrained to any particular orientation within the ab-plane of the parent structure. In this case, the ferromagnetic moment allowed by $mG_2^+$ as a secondary irrep could still be zero for thermodynamic reasons.

**Table 8** Subgroups arising from coupling of irreps $mG_5^+$ and $M_2^-$, with respect to the parent space group *P*6$_3$/*mmc*.

| SGN.M | Subgroup | OPD label $mG_5^+$ $M_2^-$ | OPD vector $mG_5^+$, $M_2^-$ | Basis Vector | Origin |
|---|---|---|---|---|---|
| 62.441 | Pnma | P1(1) P1(1) | (a,-1.732a),(b,0,0) | (0,0,1),(0,-1,0),(2,1,0) | (1/2,0,0) |
| 62.447 | Pnm'a' | P2(1) P1(1) | (a,0.577a),(b,0,0) | (0,0,1),(0,-1,0),(2,1,0) | (1/2,0,0) |
| 14.75 | P2$_1$/c | C1(1) P1(1) | (a,b),(c,0,0) | (0,1,0),(0,0,1),(2,0,0) | (1/2,0,0) |
| 64.469 | Cmca | P1(1) P2(3) | (a,-1.732a),(0,b,b) | (0,2,0),(-4,-2,0),(0,0,1) | (1,3/2,0) |
| 64.474 | Cm'c'a | P2(1) P2(3) | (a,0.577a),(0,b,b) | (0,2,0),(-4,-2,0),(0,0,1) | (1,3/2,0) |
| 36.172 | Cmc2$_1$ | P1(1) C2(3) | (a,-1.732a),(c,b,b) | (0,2,0),(-4,-2,0),(0,0,1) | (0,0,0) |
| 36.176 | Cm'c'2 | P2(1) C2(3) | (a,0.577a),(c,b,b) | (0,2,0),(-4,-2,0),(0,0,1) | (0,0,0) |
| 14.75 | P2 1/c | C1(1) C1(1) | (a,b),(c,0,d) | (0,2,0),(0,0,1),(2,0,0) | (1/2,0,0) |
| 4.7 | P2 1 | C1(1) S1(1) | (a,b),(c,d,e) | (-2,0,0),(0,0,1),(0,2,0) | (0,0,0) |

Below the spin-flop transition at ~450 K moments lie perpendicular to the ab-plane and the structure remains antiferromagnetic (Hirahara and Murakami 1958; Andresen 1960; Sparks et al 1962; Andresen and Torbo 1967; Horwood et al 1976) with spin configurations that are consistent with $mG_4^+$. Combining $mG_4^+$ with $M_2^-$ gives the list of subgroup structures in Table 9, from which *P*6$_3$'*m*'*c* would be consistent with the *P*6$_3$*mc* structure reported by Keller-Besrest and Collin (1990). *P*6$_3$'*m*'*c* is not a subgroup of *Pnma*, *Pnm*'*a*' or *P*2$_1$/*c* so the spin-flop transition is necessarily first order in character.

**Table 9** Subgroups arising from coupling of irreps $mG_4^+$ and $M_2^-$, with respect to the parent space group *P*6$_3$/*mmc*.

| SGN.M | Subgroup | OPD $mG_4^+$ $M_2^-$ | OPD vector $mG_4^+, M_2^-$ | Basis Vectors | Origin |
|---|---|---|---|---|---|
| 62.446 | Pn'm'a | P1(1) P1(1) | (a).(b,0,0) | (0,0,1),(0,-1,0),(2,1,0) | (1/2,0,0) |
| 186.205 | P6$_3$'m'c | P1(1) P3(1) | (a),(b,b,b) | (0,-2,0),(2,2,0),(0,0,1) | (0,0,0) |
| 64.476 | Cm'ca' | P1(1) P2(1) | (a),(b,0,b) | (2,0,0),(2,4,0),(0,0,1) | (1/2,-1,0) |
| 36.174 | Cm'c2$_1$' | P1(1) C2(1) | (a),(b,c,b) | (2,0,0),(2,4,0),(0,0,1) | (0,0,0) |
| 14.79 | P2$_1$'/c' | P1(1) C1(1) | (a),(b,0,c) | (0,2,0),(0,0,1),(2,0,0) | (1/2,0,0) |
| 4.9 | P2$_1$' | P1(1) S1(1) | (a),(b,c,d) | (-2,0,0),(0,0,1),(0,2,0) | (0,0,0) |

The room temperature ($P\bar{6}2c$) structure of troilite is also antiferromagnetic, with individual moments parallel to the crystallographic c-axis (Hirahara and Murakami 1958; Andresen 1960; Bertaut 1980). Combining $mG_4^+$ with H$_1$ (as identified in Section 2.6) gives $P\bar{6}'2c'$ as the most likely magnetic space group (see the first entry of Table 10).

**Table 10** Subgroups arising from coupling of irreps H$_1$ and $mG_4^+$, with respect to the parent space group $P6_3/mmc$.

| SGN.M | Space Group | OPD label | OPD vector | Basis Vector | Origin |
|---|---|---|---|---|---|
| 190.23 | P$\bar{6}$'2c' | P1(1)P1(1) | (0,0,0,a,b) | (1,-1,0),(1,2,0),(0,0,2) | (2/3,1/3,1/4) |
| 189.224 | P$\bar{6}$'2m' | P2(1)P1(1) | (0,a,0,0,b) | (1,-1,0),(1,2,0),(0,0,2) | (2/3,1/3,3/4) |
| 15.89 | C2'/c' | P4(1)P1(1) | (0,0,a,-a,b) | (-1,-2,0),(3,0,0),(0,0,2) | (1/2,0,0) |
| 12.62 | C2'/m' | P8(1)P1(1) | (a,a,0,0,b) | (-1,-2,0),(3,0,0),(0,0,2) | (1/2,0,0) |
| 174.135 | P$\bar{6}$' | C1(1)P1(1) | (0,a,0,b,c) | (2,1,0),(-1,1,0),(0,0,2) | (2/3,1/3,3/4) |
| 9.39 | Cc' | C2(1)P1(1) | (0,0,a,b,c) | (-1,-2,0),(3,0,0),(0,0,2) | (1/3,-1/3,0) |
| 8.34 | Cm' | C4(1)P1(1) | (a,b,0,0,c) | (-1,-2,0),(3,0,0),(0,0,2) | (1/3,-1/3,0) |
| 5.15 | C2' | C11(1)P1(1) | (a,-a,b,-b,c) | (-1,-2,0),(3,0,0),(0,0,2) | (0,0,1/2) |
| 5.13 | C2 | C3(1)P1(1) | (a,0,0,b,c) | (3,0,0),(1,2,0),(0,0,2) | (2/3,1/3,1/4) |
| 2.4 | P$\bar{1}$ | C10(1)P1(1) | (a,a,b,-b,c) | (0,0,2),(2,1,0),(-1,1,0) | (0,1/2,0) |
| 1.1 | P1 | 4D1(1)P1(1) | (a,b,c,d,e) | (0,0,2),(2,1,0),(-1,1,0) | (0,0,0) |

### 3.2. 4C Pyrrhotite

#### 3.2.1. Ferrimagnetism below ~590 K

If the antiferromagnetic ordering which occurs below ~590 K in FeS is essentially the same as occurs at the same temperature all across the $Fe_{1-x}S$ solid solution, it should be possible to obtain the magnetic structure of 4C pyrrhotite reported by Powell et al. (2004) by combining $mG_5^+$ with U1(1/2,0,1/4). Table 11 contains subgroups of $P6_3/mmc$ that arise by coupling of these two irreps and includes *C2'/c'* (row 10) as the most appropriate magnetic space group for the reported structure. This has direction P4 of the six component U1 order parameter, which has one non-zero component as (*a*,0,0,*a*,0,0), and direction P2 of the two component $mG_5^+$ order parameter, which has one non-zero component as (-b,0.577b). The latter gives the alignment of moments at an angle of 30° to [100] of the hexagonal parent structure, which is also the direction of the crystallographic a-axis of the monoclinic structure (Fig. 3).

Because the symmetry is reduced to monoclinic by the pattern of Fe/vacancy ordering, $mG_4^+$ is a secondary order parameter once the magnetic symmetry is broken by $mG_5^+$. This allows moments to have a component out of the ab-plane, though still constrained to be within the ac-plane of the monoclinic structure (Fig. 3). It is worth noting a similarity to behaviour observed in hematite with similar group theory underpinnings given in Harrison et al. 2010. By analogy with the magnetic behaviour of FeS, it should be expected that there will an energetic advantage in a realignment of these moments from parallel to the ab-plane at high temperatures ($mG_5^+$) to perpendicular to it at low temperatures ($mG_4^+$). In this case, all that is required is an increase in the magnitude of the $mG_4^+$ component as temperature reduces. The observed angle of rotation angle of the ferrimagnetic moment out of the ab-plane increases from ~6° at room temperature to 25-30° at ~10-30 K (Powell et al. 2004; Koulialias 2018).

**Table 11** Magnetic subgroups arising from coupling of irreps $mΓ_5^+$ and U1(1/2,0,1/4), with respect to the parent space group P6₃/mmc. Only distortions with single parameter order parameter directions have been included. Full table can be found at Table S3.

| SGN.MSGN | Subgroup | OPD U1(1/2,0,1/4) $mG_5^+$ | OPD vector U1(1/2,0,1/4), $mG_5^+$ | Basis Vectors | Origin |
|---|---|---|---|---|---|
| 12.58 | C2/m | P1(1) P1(1) | (a,0,0,0,0,0),(b,-1.732b) | (2,1,-4),(0,1,0),(2,1,0) | (0,0,0) |

| | | | | | | |
|---|---|---|---|---|---|---|
| 44.229 | Imm2 | P2(1) P1(1) | (a,-0.414a,0,0,0),(0,b,-1.732b) | (0,1,0),(0,0,4),(2,1,0) | | (0,0,1/4) |
| 12.58 | C2/m | P3(1) P1(3) | (a,0,a,0,0),(b,1.732b) | (2,-2,0),(2,2,0),(-1,1,2) | | (0,0,0) |
| 15.85 | C2/c | P4(1) P1(3) | (a,0,0,a,0,0),(b,1.732b) | (2,-2,0),(2,2,0),(-1,1,2) | | (0,1/2,0) |
| 42.219 | Fmm2 | P5(1) P1(3) | (a,-0.414a,a,-0.414a,0,0),(b,1.732b) | (2,2,0),(0,0,4),(2,-2,0) | | (1/8,-1/8,1/4) |
| 43.224 | Fdd2 | P6(1) P1(3) | (a,-0.414a,0.414a),(a,0,0,b,1.732b) | (2,2,0),(0,0,4),(2,-2,0) | | (11/8,-3/8,3/4) |
| 12.62 | C2'/m' | P1(1) P2(1) | (a,0,0,0,0),(b,0.577b) | (2,1,-4),(0,1,0),(2,1,0) | | (0,0,0) |
| 44.231 | Im'm2' | P2(1) P2(1) | (a,-0.414a,0,0,0),(b,0.577b) | (0,1,0),(0,0,4),(2,1,0) | | (0,0,1/4) |
| 12.62 | C2'/m' | P3(1) P2(3) | (a,0,a,0,0),(-b,0.577b) | (2,-2,0),(2,2,0),(-1,1,2) | | (0,0,0) |
| 15.89 | C2'/c' | P4(1) P2(3) | (a,0,0,a,0,0),(-b,0.577b) | (2,-2,0),(2,2,0),(-1,1,2) | | (0,1/2,0) |
| 42.221 | Fm'm2' | P5(1) P2(3) | (a,-0.414a,a,-0.414a,0,0),(-b,0.577b) | (2,2,0),(0,0,4),(2,-2,0) | | (1/8,-1/8,1/4) |
| 43.226 | Fd'd2' | P6(1) P2(3) | (a,-0.414a,0.414a,a,0,0),(-b,0.577b) | (2,2,0),(0,0,4),(2,-2,0) | | (11/8,-3/8,3/4) |

Table 12 contains a full list of the magnetic Γ-point irreps which are allowed by symmetry to have non-zero values in the *C*2'/*c*' structure. In principle, $mG_2^+$ (*a*) and $mG_6^+$ (*a*,1.732*a*) allow ferromagnetic moments parallel to the crystallographic c*- and b-axes of the monoclinic structure, respectively. These have not been observed and must be effectively zero for reasons of thermodynamic stability.

**Table 12** Non-zero irreps arising from the symmetry changes *P*6$_3$/*mmc* → *C*2'/*c*' and *P*6$_3$/*mmc* → $P\bar{1}$.

| | U1(1/2,0,1/4) | $mG_2^+$ | $mG_4^+$ | $mG_5^+$ | $mG_6^+$ |
|---|---|---|---|---|---|
| *C*2'/*c*' | P4 (a,0,0,a,0,0) | P1 (a) | P1 (a) | P2 (-a,0.577a) | P1 (a,1.732a) |
| $P\bar{1}$ | C5 (a,0,0,b,0,0) | P1 (a) | P1 (a) | C1 (a,b) | C1 (a,b) |

**3.2.2. Besnus transition in 4C Pyrrhotite**

The Besnus transition near 35 K in 4C pyrrhotite (~$Fe_7S_8$) is associated with changes in magnetic properties (Fillion and Rochette 1988; Dekkers et al. 1989; Rochette et al. 1990; Kind et al. 2013; Baranov et al. 2015; Koulialias et al. 2016; Volk et al. 2016; Bazaeva et al. 2016; Koulialias et al 2018; Horng and Roberts 2018), an anomaly in heat capacity (Grønvold et al. 1959), an anomaly in resistivity (Besnus and Meyer 1964; Charilaou et al. 2015), apparently continuous changes in the temperature dependence of lattice parameters (Fillion et al. 1992; Koulialias et al. 2018), together with small adjustments of Fe-Fe bond lengths and Fe-S-Fe bond angles (Koulialias et al. 2018). The evidence from Mössbauer spectroscopy is that the transition is marked by splitting of at least one of the Fe sites into two equally populated subsites (Jeandey et al. 1991; Oddou et al. 1992; Ferrow et al. 2006). In combination, these features are consistent with a phase transition between structures with a group-subgroup relationship. P1 and $P\bar{1}$ have been suggested for the crystallographic space group of the low temperature structure (Wolfers et al. 2011).

The pattern of vacancy ordering must remain the same through the transition because of the slowness of Fe/vacancy diffusion at low temperatures, which leaves $P\bar{1}$ as the only possible subgroup of $C2'/c'$ in Table S3. This has order parameter direction C5, ($a$,0,0,$b$,0,0), for U1(1/2,0,1/4), C1 ($c$,$d$) for $mG_5^+$ and the additional secondary magnetic Γ-point irreps listed in Table 10. Non-zero values of $mG_2^+$ and $mG_6^+$ allow ferromagnetic moments parallel to the crystallographic c* and b-axes of the monoclinic structure but there is no experimental evidence suggesting that these develop. In terms of the magnetic structure alone, the transition can be understood as a rotation of the direction of individual moments out of the ac-plane of the monoclinic structure. There are no symmetry constraints on the orientation of antiferromagnetic/ferrimagnetic moments in the $P\bar{1}$ structure but, because of the slowness of Fe/vacancy reordering, the order parameter will effectively remain ($a$,0,0,$a$,0,0) instead of relaxing to ($a$,0,0,$b$,0,0). If this influences the magnetism, it is possible that the $mG_5^+$ order parameter could remain close to ($b$,$b$), which would fix the orientation of the component in the ab-plane as being at an angle of 45° from the [100] and [120] directions of the parent hexagonal structure. Experimental data for magnetic properties below the transition point are consistent with a change of easy axis within the ab-plane (Rochette et al. 1990; Wolfers et al. 2011; Volk et al 2016; Koulialias et al. 2018). Wolfers et al. (2011) found two easy directions, close to 45° and 135° but presumably there would only be one of these in an untwinned crystal.

This symmetry analysis does not depend on identification of the driving mechanism for the transition, which might be due primarily to a change in electronic structure (e.g., Besnus and Meyer 1964; Fillion and Rochette 1988; Ferrow et al. 2006; Koulialias et al. 2016; Koulialias et al. 2018) or, more simply, to the normal interactions between spins which give rise to changes in the easy axis of magnetisation.

We summarise the sequence of spin orientations in Fig. 5.

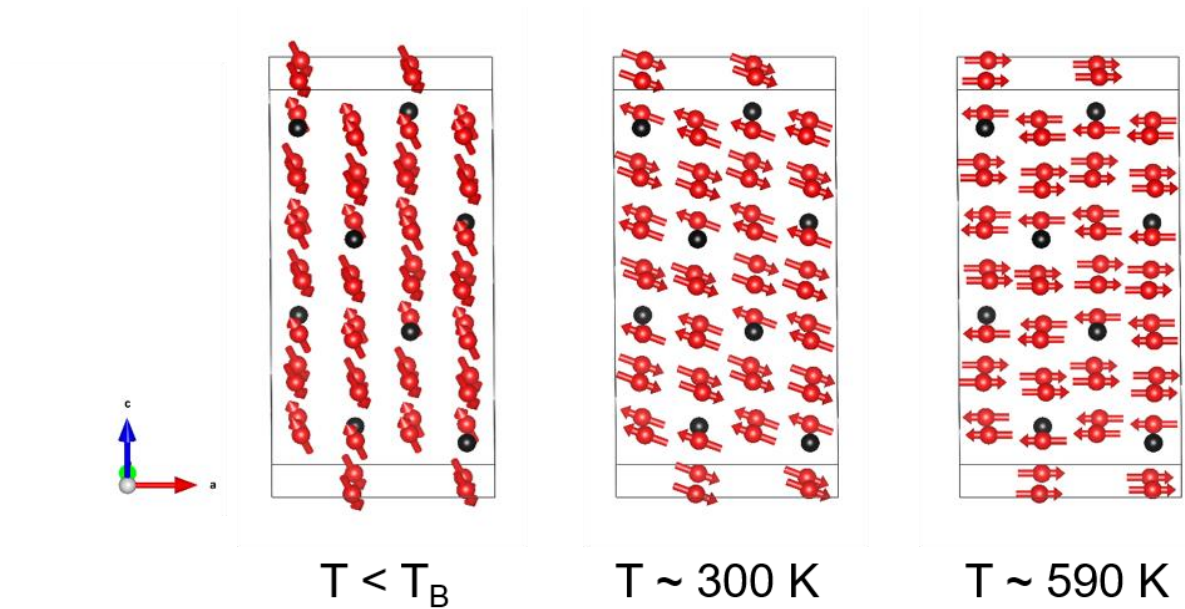

**Figure 5** Temperature dependence of the spin orientation in 4C pyrrhotite. Sulphur atoms have been removed for clarity. Iron atoms are red with red arrows for the moment direction. The vacant sites are shown in black.

**3.3. Magnetism in 3C pyrrhotite**

The 3C structure of pyrrhotite in space group $P3_121$ is similar to that of the 4C structure in the sense that vacancies are confined to alternate Fe-layers. There appears to be little information in the literature for its magnetic structure, but the equivalent phase in $Fe_7Se_8$ is ferrimagnetic (Kawaminami and Okazaki 1970), as would be expected for this distribution of vacancies. Horng and Roberts (2018) present magnetic data on 3C pyrrhotite that suggests a remanence of the same order as 4C. This would make sense on the basis of our group theory results as the remanence will come from the vacancies, of which there are the same concentration in both structures. Furthermore, it is interesting to note that in all of their samples (which are all multiphase) there are none with both 3C and 4C. They do not observe a transition in 3C pyrrhotite in the range 5 – 300 K.

Combinations of U1(1/2,0,1/3) with $mG_5^+$ and $mG_4^+$ yield the possible magnetic space groups listed in Tables 13 and 14.

**Table 13** Magnetic subgroups arising from coupling of irreps $m\Gamma_5^+$ and U1(1/2,0,1/3), with respect to the parent space group $P6_3/mmc$. Only distortions with single parameter order parameter directions have been included. Full table can be found at Table S12.

| SGN.MSGN | Subgroup | OPD | OPD vector | Basis Vectors | Origin |
|---|---|---|---|---|---|
| 58.393 | Pnnm | P1(1)P1(1) | (a,0,0,0,0,b,-1.732b) | (-2,-1,0),(0,0,3),(0,1,0) | (0,0,0) |
| 59.405 | Pmmn | P2(1)P1(1) | (0,a,0,0,0,b,-1.732b) | (0,1,0),(0,0,3),(2,1,0) | (1/2,0,0) |
| 64.469 | Cmca | P5(1)P1(2) | (a,0,0,0,a,0,-2b,0) | (2,0,0),(2,4,0),(0,0,3) | (0,0,0) |
| 63.457 | Cmcm | P6(1)P1(2) | (0,a,0,0,0,-a,-2b,0) | (2,0,0),(2,4,0),(0,0,3) | (1/2,0,0) |
| 58.398 | Pnn'm' | P1(1)P2(1) | (a,0,0,0,0,b,0.577b) | (0,0,3),(2,1,0),(0,1,0) | (0,0,0) |
| 59.41 | Pmm'n' | P2(1)P2(1) | (0,a,0,0,0,b,0.577b) | (0,0,3),(0,-1,0),(2,1,0) | (1/2,1/2,0) |
| 64.474 | Cm'c'a | P5(1)P2(2) | (a,0,0,0,a,0,0,-1.155b) | (2,0,0),(2,4,0),(0,0,3) | (0,0,0) |
| 63.462 | Cm'c'm | P6(1)P2(2) | (0,a,0,0,0,-a,0,-1.155b) | (2,0,0),(2,4,0),(0,0,3) | (1/2,0,0) |

**Table 14** Magnetic subgroups arising from coupling of irreps m$\Gamma_4^+$ and U1(1/2,0,1/3), with respect to the parent space group *P6$_3$/mmc*. Only distortions with single parameter order parameter directions have been included. Full table can be found at Table S11

| SGN.MSGN | Subgroup | OPD | OPD vector | Basis Vectors | Origin |
|---|---|---|---|---|---|
| 58.398 | Pnn'm' | P1(1)P1(1) | (a,0,0,0,0,b) | (-2,-1,0),(0,0,3),(0,1,0) | (0,0,0) |
| 59.409 | Pm'm'n | P2(1)P1(1) | (0,a,0,0,0,b) | (0,0,3),(0,-1,0),(2,1,0) | (1/2,1/2,0) |
| 164.89 | P$\bar{3}$m'1 | P3(1)P1(1) | (a,0,a,0,a,0,b) | (0,-2,0),(2,2,0),(0,0,3) | (0,0,0) |
| 187.211 | P$\bar{6}$'m'2 | P4(1)P1(1) | (a,-0.577a,a,-0.577a,a,-0.577a,b) | (2,2,0),(-2,0,0),(0,0,3) | (0,0,1/4) |
| 64.476 | Cm'ca' | P5(1)P1(1) | (a,0,0,0,a,0,b) | (2,0,0),(2,4,0),(0,0,3) | (0,0,0) |
| 63.464 | Cm'cm' | P6(1)P1(1) | (0,a,0,0,0,-a,b) | (2,0,0),(2,4,0),(0,0,3) | (1/2,0,0) |
| 152.35 | P3$_1$2'1 | P7(1)P1(1) | (a,1.732a,-2a,0,a,-1.732a,b) | (0,-2,0),(2,2,0),(0,0,3) | (0,0,0) |
| 151.29 | P3$_1$12 | P8(1)P1(1) | (a,0.577a,-a,0.577a,0,-1.155a,b) | (2,2,0),(-2,0,0),(0,0,3) | (0,0,1/4) |

Magnetic moments within the *ab*-plane would be produced by m$G_5^+$, giving one of the structures seen in Table 13 as the most likely symmetry immediately below ~590 K. The most likely magnetic structure with moments perpendicular to the ab-plane would have space group P3$_1$2'1 (Table 14). Unlike the case of the monoclinic 4C structure, m$G_4^+$ does not become a secondary irrep when the symmetry is broken by m$G_5^+$. A change in the orientation of moments with falling temperature from parallel to perpendicular to the ab-plane would therefore be expected to occur by a discrete spin-flop transition. This occurs at ~130K in 3C $Fe_7Se_8$ (Andresen and Leciejewicz 1964; Kawaminami and Okazaki 1970; Ericsson et al. 1997).

**3.4. Magnetism in 5C**

The crystallographic space groups of the structures of pyrrhotite with 5C and 6C superstructures are perhaps less certain than those of the better-understood 3C and 4C structures. Both are reported to have partially ordered Fe/vacancy sites and monoclinic symmetry: *Cc* for the 6C structure (Koto et al. 1975; de Villiers and Liles 2010) and $P2_1$ as the most recent proposal for the 5C structure (Liles and de Villiers 2012). The expected pattern with respect to their magnetic properties would be antiferromagnetic (or ferrimagnetic, depending on the distribution of vacancies), with moments in the ab-plane at high temperatures and perpendicular to the ab-plane at low temperatures. Whether the change in orientation during cooling occurs by a continuous rotation or abruptly at a spin-flop transition depends on whether the combination of the U1 order parameter for Fe/vacancy ordering and $mG_5^+$ for magnetic ordering gives $mG_4^+$ as a secondary order parameter. U1(1/2,0,1/5) combined with $mG_5^+$ does not. U1(1/2,0,1/6) combined with $mG_5^+$ does. The energetically preferred orientation of moments within the ab-plane will be described, as for the other superstructure types, by P1, P2 or C1 directions of irrep $mG_5^+$.

In none of the reported 5C structures (nor indeed in any but the triclinic structures) is it possible to have both $m\Gamma_4^+$ and $m\Gamma_5^+$ allowed together. Therefore, it seems likely that 5C pyrrhotite will show an abrupt spin flop transition from having spins in the *ab*-plane to having the spins perpendicular to that plane. The reported structures are all compatible with coupling of the vacancy ordering to either $m\Gamma_4^+$ or $m\Gamma_5^+$. The most likely scenario would be a phase transition from one of the higher symmetry structures, Cmca or $P2_1/c$ to the $P2_1$ structure reported by Liles et al. (2012) predominantly using data collected at 120 K. The higher symmetry structure would be due to a coupling of the U1 irrep with the $m\Gamma_5^+$ irrep and Table 15 shows the possible distortions. The third entry shows the possible Cmca structure whilst the $P2_1/c$ structure is seen in entries 5, 8 and 10. The lower temperature, low symmetry structure would arise from a coupling of the U1 irrep with the $m\Gamma_4^+$ irrep (see final entry of Table 16).

**Table 15** Magnetic subgroups arising from coupling of irreps $m\Gamma_5^+$ and U1(1/2,0,1/5), with respect to the parent space group $P6_3/mmc$. Some entries have been omitted for clarity and brevity. Full table can be found at Table S6.

| SGN.m | Space group | OPD Name | OPD vector | Basis Vectors | Origin |
| --- | --- | --- | --- | --- | --- |

| SGN.m | Space group | OPD Name | OPD vector | Basis Vectors | Origin |
|---|---|---|---|---|---|
| 58.393 | Pnnm | P1(1)P1(1) | (a,0,0,0,0,b,-1.732b) | (-2,-1,0),(0,0,5),(0,1,0) | (0,0,0) |
| 59.405 | Pmmn | P2(1)P1(1) | (0,a,0,0,0,b,-1.732b) | (0,1,0),(0,0,5),(2,1,0) | (1/2,0,0) |
| 64.469 | Cmca | P5(1)P1(2) | (a,0,0,0,a,0,-2b,0) | (2,0,0),(2,4,0),(0,0,5) | (0,0,0) |
| 63.457 | Cmcm | P6(1)P1(2) | (0,a,0,0,0,-a,-2b,0) | (2,0,0),(2,4,0),(0,0,5) | (1/2,0,0) |
| 14.75 | P2$_1$/c | P1(1)C1(1) | (a,0,0,0,0,0,b,c) | (0,1,0),(0,0,5),(2,0,0) | (0,0,0) |
| 11.5 | P2$_1$/m | P2(1)C1(1) | (0,a,0,0,0,0,b,c) | (-2,0,0),(0,0,5),(0,1,0) | (-1/2,0,0) |
| 4.7 | P2$_1$ | C1(1)C1(1) | (a,b,0,0,0,0,c,d) | (-2,0,0),(0,0,5),(0,1,0) | (-1/2,0,0) |
| 14.75 | P2$_1$/c | C4(1)C1(1) | (0,0,a,0,b,0,c,d) | (-2,0,0),(0,0,5),(0,2,0) | (0,0,0) |
| 11.5 | P2$_1$/m | C5(1)C1(1) | (0,0,0,a,0,b,c,d) | (-2,0,0),(0,0,5),(0,2,0) | (0,1/2,0) |
| 14.75 | P2$_1$/c | C6(1)C1(1) | (a,0,0,0,0,b,c,d) | (-2,0,0),(0,0,5),(2,2,0) | (0,1/2,0) |
| 4.7 | P2$_1$ | 4D1(1)C1(1) | (0,0,a,b,c,d,e,f) | (-2,0,0),(0,0,5),(0,2,0) | (0,1/2,0) |

**Table 16** Magnetic subgroups arising from coupling of irreps m$\Gamma_4^+$ and U1(1/2,0,1/5), with respect to the parent space group *P*6$_3$/*mmc*. Some entries have been omitted for clarity and brevity. Full table can be found at Table S5.

| SGN.m | Space group | OPD Name | OPD vector | Basis Vectors | Origin |
|---|---|---|---|---|---|
| 58.398 | Pnn'm' | P1(1)P1(1) | (a,0,0,0,0,0,b) | (-2,-1,0),(0,0,5),(0,1,0) | (0,0,0) |
| 59.409 | Pm'm'n | P2(1)P1(1) | (0,a,0,0,0,0,b) | (0,0,5),(0,-1,0),(2,1,0) | (1/2,1/2,0) |
| 164.89 | P$\bar{3}$m'1 | P3(1)P1(1) | (a,0,a,0,a,0,b) | (0,-2,0),(2,2,0),(0,0,5) | (0,0,0) |
| 187.211 | P$\bar{6}$'m'2 | P4(1)P1(1) | (a,-0.325a,a,-0.325a,a,-0.325a,b) | (2,2,0),(-2,0,0),(0,0,5) | (0,0,1/4) |
| 64.476 | Cm'ca' | P5(1)P1(1) | (a,0,0,0,a,0,b) | (2,0,0),(2,4,0),(0,0,5) | (0,0,0) |
| 63.464 | Cm'cm' | P6(1)P1(1) | (0,a,0,0,0,-a,b) | (2,0,0),(2,4,0),(0,0,5) | (1/2,0,0) |
| Entries removed. Selected higher order OPD entries listed below | | | | | |
| 14.79 | P2$_1$'/c' | C4(1)P1(1) | (0,0,a,0,b,0,c) | (-2,0,0),(0,0,5),(0,2,0) | (0,0,0) |
| 11.54 | P2$_1$'/m' | C5(1)P1(1) | (0,0,0,a,0,b,c) | (-2,0,0),(0,0,5),(0,2,0) | (0,1/2,0) |
| 14.79 | P2$_1$'/c' | C6(1)P1(1) | (a,0,0,0,0,b,c) | (-2,0,0),(0,0,5),(2,2,0) | (0,1/2,0) |
| 4.9 | P2$_1$' | 4D1(1)P1(1) | (0,0,a,b,c,d,e) | (-2,0,0),(0,0,5),(0,2,0) | (0,1/2,0) |

### 3.5. Magnetism in 6C

The magnetic phase diagram for 6C pyrrhotite could follow very closely that observed in 4C pyrrhotite. The C2'/c' structure attained from the coupled vacancy ordering (the P4 direction of the U1 distortion) and magnetism (the P2 direction of the m$\Gamma_5^+$ distortion), line 10 of Table 17, can then have a transition to the same P$\bar{1}$ structure, last entry Table 17, seen below the Besnus transition in 4C pyrrhotite. Magnetically this is identical. Crucially, the m$\Gamma_4^+$ irrep is a secondary distortion. This means that the moments are allowed by the symmetry to rotate out of the ab-plane. Indeed, similarly to 4C pyrrhotite, the moments can in principle line up parallel to the c-axis of the parent structure. Therefore, there is no need for an abrupt Morin-type transition and a transition like the Besnus transition is expected. The moments, which are confined to the *ac*-plane above the transition, are free to rotate out of it below the transition. However due to the nature of the vacancy ordering, a vacancy-containing bi-layer separated by a non-vacancy containing layer, the vacancies are evenly distributed on both of the anti-ferromagnetic sublattices and the overall structure is anti-ferromagnetic above and below the low-temperature transition.

**Table 17** Magnetic subgroups arising from coupling of irreps $mG_5^+$ and U1(1/2,0,1/6), with respect to the parent space group *P*6$_3$/*mmc*. Some entries have been omitted for clarity and brevity. Full table can be found at Table S9.

## 4. Discussion and conclusions

| Space Group Number | Space group | OPD Name | OPD Vector | Basis Vectors | Origin |
|---|---|---|---|---|---|
| 12.58 | C2/m | P1(1)P1(1) | (a,0,0,0,0,0,b,-1.732b) | (2,1,-6),(0,1,0),(2,1,0) | (0,0,0) |
| 44.229 | Imm2 | P2(1)P1(1) | (a,-0.268a,0,0,0,0,b,-1.732b) | (0,1,0),(0,0,6),(2,1,0) | (0,0,1/4) |
| 12.58 | C2/m | P3(1)P1(3) | (a,0,a,0,0,0,b,1.732b) | (2,-2,0),(2,2,0),(-1,1,3) | (0,0,0) |
| 15.85 | C2/c | P4(1)P1(3) | (a,0,0,a,0,0,b,1.732b) | (2,2,0),(-2,2,0),(1,1,3) | (0,1/2,0) |
| 43.224 | Fdd2 | P5(1)P1(3) | (a,-0.268a,0.268a,a,0,0,b,1.732b) | (-2,2,0),(0,0,6),(2,2,0) | (5/6,5/6,5/2) |
| 42.219 | Fmm2 | P6(1)P1(3) | (a,-0.268a,a,-0.268a,0,0,b,1.732b) | (2,2,0),(0,0,6),(2,-2,0) | (1/12,-1/12,1/4) |
| 12.62 | C2'/m' | P1(1)P2(1) | (a,0,0,0,0,0,b,0.577b) | (2,1,-6),(0,1,0),(2,1,0) | (0,0,0) |
| 44.231 | Im'm2' | P2(1)P2(1) | (a,-0.268a,0,0,0,0,b,0.577b) | (0,1,0),(0,0,6),(2,1,0) | (0,0,1/4) |
| 12.62 | C2'/m' | P3(1)P2(3) | (a,0,a,0,0,0,-b,0.577b) | (2,-2,0),(2,2,0),(-1,1,3) | (0,0,0) |
| 15.89 | C2'/c' | P4(1)P2(3) | (a,0,0,a,0,0,-b,0.577b) | (2,2,0),(-2,2,0),(1,1,3) | (0,1/2,0) |
| 43.226 | Fd'd2' | P5(1)P2(3) | (a,-0.268a,0.268a,a,0,0,-b,0.577b) | (-2,2,0),(0,0,6),(2,2,0) | (5/6,5/6,5/2) |
| 42.221 | Fm'm2' | P6(1)P2(3) | (a,-0.268a,a,-0.268a,0,0,-b,0.577b) | (2,2,0),(0,0,6),(2,-2,0) | (1/12,-1/12,1/4) |
| Entries removed | | | | | |
| 2.4 | P$\bar{1}$ | C5(1)C1(1) | (a,0,0,b,0,0,c,d) | (1,1,3),(0,2,0),(-2,0,0) | (0,1/2,0) |
| Entries removed | | | | | |

### 4.1. Fe/vacancy ordering

From the perspective of symmetry, the apparent diversity of pyrrhotite structures observed across the solid solution FeS -~Fe$_7$S$_8$ fits with the relatively simple pattern of structural and magnetic behaviour summarised schematically in Figure 6. *P*6$_3$*mc* and *Pnma* structures of FeS represent one end-member, and the series as a whole can be thought of as a combination of the M-point instability, $M_2^-$ $k$ = (1/2,0,0), combined with vacancy ordering to give a repeat along $c^*$. Divergence from the M-point, corresponding to increasing values of $g$ in irreps of the form U1(1/2,0,$g$) along the line between the M- and L-points, occurs with increasing vacancy concentration. Commensurate ordering schemes develop at particular stoichiometries with values of $g$ = 1/6, 1/5, 1/4 or 1/3. The precise distribution of vacancies in each case can be described using ratios of the amplitudes of selected secondary irreps. Incommensurate ordering schemes, defined as those with irrational values of 1/$g$, appear to be favoured at high temperatures and in crystals with compositions which fall between the nominally rational stoichiometries. Crystals with compositions in between the ideal stoichiometries for commensurate Fe/vacancy ordering schemes are expected to develop exsolution textures and/or stacking mistakes in the ab-plane layers.

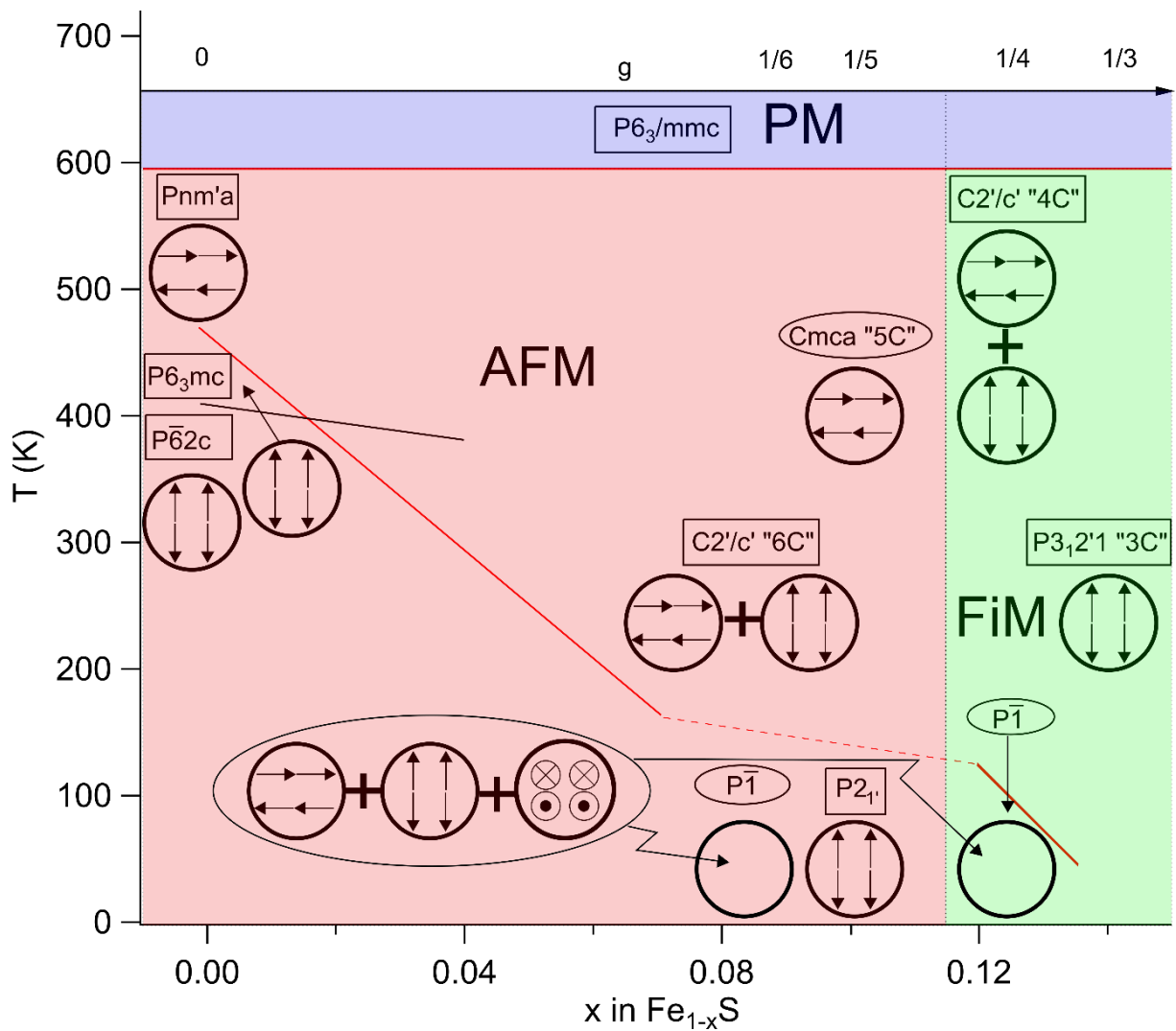

**Figure 6** Schematic phase relationships for different structures in the Fe$_{1-x}$S solid solution as obtained from a group theoretical analysis. The pink area is antiferromagnetic (AFM), the blue is paramagnetic (PM) and the green is ferrimagnetic (FiM). Space groups in rectangles are those known from scattering experiments (references given previously) known and are placed at approximately the temperature that the data used in the refinement was collected. Space groups in ellipses are those we predict based on our group theory framework. Magnetic structure is given by three cartoons representing the three different magnetic distortions: the antiparallel vertical arrows represent the m$\Gamma_4^+$ distortion, the antiparallel horizontal arrows represent the P2 direction of the m$\Gamma_5^+$ distortion and the two in and two out of page symbols represent the P1 direction of the m$\Gamma_5^+$ distortion. The horizontal red line at 595 K is the Néel temperature. The red line on the low vacancy concentration is the spin flop transition according to Horwood et. al. (1976). The black line is the structural phase transition associated with an H-point instability. The dashed red line shows the approximate temperatures of the predicted magnetically driven phase transitions in 5C and 6C pyrrhotite.

According to Ricci and Bousquet (2016), there are other possible structural instabilities in this system, but the only one which occurs at low pressures is associated with the H-point, rather than the M-point, of the Brillouin zone, giving the $P\bar{6}2c$ structure of FeS. The FeS phase diagram with respect to pressure and temperature displays considerable phenomenological richness (Fei et al 1995; Terranova and de Leeuw 2017), and the additional possibilities might perhaps be contributing factors to this.

### 4.2. Magnetic ordering

The overall pattern of magnetic ordering can also be rationalised in a relatively simple manner. Below the Néel point of ~590 K, the stable magnetic structures have moments lined up within the ab-plane of the parent hexagonal structure, as represented by the magnetic irrep, $mG_5^+$. The orientation of moments within this plane generally remains to be determined but there are three possibilities, represented by the order parameter directions P1, P2 and C1 listed in Table 7. The favoured direction in the 4C structure is P2, in which the moments are parallel to the a-axis of the monoclinic cell (or the [120] direction of the parent). The magnetic space groups shown in Figure 6 are based on the assumption that this might also be preferred in structures with different Fe/vacancy ordering schemes. At low temperatures, the alignment of magnetic moments is perpendicular to the ab-plane, as represented by the magnetic irrep, m$\Gamma_4^+$, which has only one direction, P1.

With falling temperature there will be a Morin-type spin-flop transition in structures which do not have m$\Gamma_4^+$ as a secondary order parameter resulting from the combination of U1(1/2,0,$g$) and m$\Gamma_5^+$. Two examples of this are the transition near 450 K in FeS and near 130 K in 3C Fe$_7$Se$_8$. In structures where the symmetry is broken by vacancy and magnetic ordering, such that m$\Gamma_4^+$ is a secondary irrep, rotation of the moments out of the ab-plane can occur continuously, as is observed in 4C Fe$_7$S$_8$. The subsequent Besnus transition can be understood in terms of a symmetry change $C2'/c' - P\bar{1}$, arising by the change from direction P2 to C1 of irrep $m\Gamma_5^+$ when the moments rotate out of the ac-plane of the monoclinic cell. The 6C structure bears a strong similarity to that seen in the 4C case. We expect to see a similar continuous rotation of the spins out of the *ac*-plane. 6C has been reported in the Cc structure but inspection of the original determination suggests that the true space group is C2/c. Therefore, we predict that the magnetic structure is C2'/c' and the low temperature transition will be to P$\bar{1}$ in the same way as for the Besnus transition in the case of 4C pyrrhotite. We would also predict that the transition temperature will fall between 170 K (seen for off-stoichiometry FeS) and 35 K (the Besnus transition).

The magnetic properties of 5C pyrrhotite will be most closely related to those seen in the 3C case, with an abrupt spin flop transition. The 5C structure is less conclusively determined. Cmca (Cmce in new notation), P$2_1$/c and P$2_1$ have all been reported. All of these possibilities appear in the table in the appendix for the distortions arising from coupled U1 and $m\Gamma_5^+$ irreps. Cmca and P$2_1$/c arise from a single order parameter distortion and P$2_1$ from a higher order distortion. We therefore suggest that there is a transition from one of Cmca or P$2_1$/c to P$2_1$. We would then expect this transition to occur at a lower temperature than the transition in 6C pyrrhotite but above the temperature of the Besnus transition (35 K).

**4.3. Coupling between Fe/vacancy and magnetic ordering**

The overall superposition of structure types in Figure 6 allows interactions between Fe/vacancy and magnetic ordering. In this context, the key issue is the form and strength of coupling between order parameters with symmetries determined by U1(1/2,0,$g$), $m\Gamma_4^+$ and $m\Gamma_5^+$. If the coupling is expressed in terms of two order parameters, $Q_V$ to represent vacancy ordering and $M_{G5+}$ to represent magnetic ordering on the basis of $m\Gamma_5^+$, the lowest order term permitted by symmetry is biquadratic, $\lambda_{\Gamma5+} Q_V^2 M_{\Gamma5+}^2$. The total excess free energy is given by $G_V + G_M + \lambda_{\Gamma5+} Q_V^2 M_{\Gamma5+}^2$, where $G_V$ is the contribution from vacancy ordering alone and $G_M$ is due to the magnetisation alone. It is notable that the Néel temperature does not vary significantly with composition across the solid solution. There also does not appear to be any correlation between magnetisation on the basis of $m\text{G}_5^+$ and the distribution of phase boundaries for the different structure types at high temperatures in the phase diagram (Fig. 1). As a first approximation, therefore, it would appear that the evolution of $M_{G5+}$ is overwhelmingly dominated by $G_M$, ie by interactions between spins in the classic manner for a paramagnetic – antiferromagnetic/ferromagnetic transition.

By way of contrast with the behaviour at the Néel point, subsequent changes in magnetic structure differ between structure types and thus appear to be sensitive to the precise distribution of vacancies. This has been seen, for example, as a change of transition temperature for an antiferromagnetic – ferromagnetic transition in a hexagonal sample with composition Fe$_9$S$_{10}$ due to changes in the degree of vacancy order (Bennett and Graham 1981). The driving energy for a spin-flop transition or a continuous rotation of moments out of the ab-plane is the difference between $G_V + G_M + \lambda_{\Gamma 5+} Q_V^2 M_{\Gamma 5+}^2$, and $G_V + G_M + \lambda_{\Gamma 5+} Q_V^2 M_{\Gamma 4+}^2$. Even if small, the coupling energies can presumably provide a subtle influence on the overall thermodynamic stability of different magnetically ordered structures.

The only possibility for linear-quadratic coupling is associated with the Besnus transition. If it is driven simply by interactions between spins, the coupling is still biquadratic and the monoclinic – triclinic transition will be improper ferroelastic. If, on the other hand, the driving mechanism is a change in electronic structure, a new order parameter, $Q_E$, would need to be specified. This will have the symmetry of a zone centre irrep, the transition will be pseudo-proper ferroelastic and the coupling can be of the form $\lambda Q_E M^2$.

**4.4. Coupling with strain**

An important mechanism for coupling between two order parameters is by overlap of the strains associated with one order parameter with the strains associate with the other. Inspection of the lattice parameters given by Powell et al (2004) for a 4C sample and by Li and Frantzen (1996) for FeS show changes in linear strains on the order of ~1-3% associated with the transition at ~590 K. No formal analysis of these has yet been undertaken, so the separate contributions of coupling with structural, Fe/vacancy and magnetic order parameters remain to be distinguished. In anticipation of such an analysis, selected Landau expansions which include the lowest order coupling terms allowed by symmetry have been set out in Appendix B. Strain coupling with the spin-flop transition at ~450 K and the Besnus transition near 35 K in $Fe_7S_8$ is very much weaker, based on the very much smaller changes in lattice parameters given by Li and Frantzen (1996), Fillion et al. (1992) and Koulialias et al. (2018)

Strain coupling effects are also detectable through their influence on elastic constants. For example, an experimental method for distinguishing between linear-quadratic and biquadratic coupling at the Besnus transition, would be to observe the temperature dependence of single-crystal elastic constants. A transition in which the monoclinic symmetry is broken by a structural order parameter, $Q$, could have coupling with the symmetry-breaking shear strain, $e_{sb}$, as $\lambda e_{sb} Q$. If the symmetry is broken by a magnetic order parameter, $M$, the coupling would be $\lambda e_{sb} M^2$. A transition driven by $Q$ would therefore be pseudoproper ferroelastic while a transition driven by $M$ would be improper ferroelastic, and the difference would give rise to quite distinct patterns of evolution for elastic constants $C_{44}$ and $C_{66}$ (see, for example, Figure 4 of Carpenter and Salje 1998). Even the most recent diffraction measurements (Koulialias et al. 2018) have not yet resolved a distortion from monoclinic lattice geometry, however, implying that coupling with $e_{sb}$ is weak.

### 4.5. Multiferroicity

As an ongoing matter of interest with respect to possible multiferroic sulphides, it is worth noting that structures with $P\bar{6}2c$ (FeS at room temperature) and $P3_121$ (3C pyrrhotite at room temperature) symmetry are expected to be piezoelectric while those with $P6_3mc$ (FeS between ~420 K and 460 K) or $P2_1$ (5C pyrrhotite at low temperature) symmetry could be ferroelectric.

### Appendix A. Examples of input/output information for ISOTROPY.

Figure A1 is an example of the output from ISOTROPY showing a list of possible secondary irreps when the space group of the parent structure is *P*6$_3$/*mmc* (no. 194) and that of the product structure is *C*2/*c* (no. 15) with the unit cell of 4C pyrrhotite, as specified by the choice of basis vectors and origin. These choices are indicated by the letter v at the start of the line. The list of secondary irreps shows the subgroup which would develop if each of them acted alone and their size.

The first step in the process is to compile a list of secondary distortions. When we know the starting and finishing space groups we can use ISOTROPY to determine the irreps, along with the pertinent OPDs.

```
Isotropy, Version 9.4.1, December 2015
Harold T. Stokes, Dorian M. Hatch, and Branton J. Campbell
Brigham Young University
*DISPLAY SETTING
Current setting is International (new ed.) with conventional basis vectors.
*DISPLAY SETTING IRREP
Current irrep version is 2011
Use "VALUE IRREP VERSION" to change version
*v par 195
*v sub 15
*v bas 2,-2,0 2,2,0 -1,1,2
*v ori 0,1/2,0
*d dir
Error: not all elements of the subgroup are elements of parent group
*v par 194
*v sub 15
*v bas 2,-2,0 2,2,0 -1,1,2
*v ori 0,1/2,0
*d dir
Irrep  (ML) k params Dir                   Subgroup       Size
GM1+                 (a)                   194 P6_3/mmc   1
GM3+                 (a)                   164 P-3m1      1
GM5+                 (a,1.732a)            63  Cmcm       1
GM6+                 (a,-0.577a)           12  C2/m       1
A2                   (a,a)                 165 P-3c1      2
A3                   (a,0.577a,-0.577a,a)  15  C2/c       2
L1                   (0,0,0,0,a,a)         12  C2/m       2
M1-                  (0,0,a)               52  Pnna       2
M3-                  (0,0,a)               57  Pbcm       2
U1          1/4      (a,0,0,a,0,0)         15  C2/c       8
U2          1/4      (0,a,a,0,0,0)         15  C2/c       8
U3          1/4      (a,0,0,a,0,0)         15  C2/c       8
U4          1/4      (0,a,a,0,0,0)         15  C2/c       8
```

**Figure 7** Using ISOTROPY to determine the irreps responsible for a distortion characterised by a known parent, subgroup and unit cell.

In order to determine which irreps can influence occupancy we specify the relevant Wyckoff site (*a*) and use ISOTROPY to give the effect of the irrep on the occupancy by using show microscopic scalar (sh mic scal). This is used as the occupancy is a scalar.

```
Isotropy, Version 9.4.1, December 2015
Harold T. Stokes, Dorian M. Hatch, and Branton J. Campbell
Brigham Young University
*DISPLAY SETTING
Current setting is International (new ed.) with conventional basis vectors.
*DISPLAY SETTING IRREP
Current irrep version is 2011
Use "VALUE IRREP VERSION" to change version
*v par 194
*v sub 15
*v irrep u1
*v kv 1,1/4
*v dir p4
*v bas 2,-2,0 2,2,0 -1,1,2
*v ori 0,1/2,0
*v wy a
*sh par
*sh sub
*sh mic sc
*v cell 2,0,0 0,2,0 0,0,4
*d dis
Parent          Point       Projected Order Functions
194 P6_3/mmc    (0,0,0)     f
                (0,0,1)     -f
                (0,0,2)     -f
                (0,0,3)     f
                (0,1,0)     f
                (0,1,1)     f
                (0,1,2)     -f
                (0,1,3)     -f
                (1,0,0)     -f
                (1,0,1)     -f
                (1,0,2)     f
                (1,0,3)     f
                (1,1,0)     -f
                (1,1,1)     f
                (1,1,2)     f
                (1,1,3)     -f
                (0,0,1/2)   0
                (0,0,3/2)   -1.414f
                (0,0,5/2)   0
                (0,0,7/2)   1.414f
                (0,1,1/2)   1.414f
                (0,1,3/2)   0
                (0,1,5/2)   -1.414f
                (0,1,7/2)   0
                (1,0,1/2)   -1.414f
                (1,0,3/2)   0
                (1,0,5/2)   1.414f
                (1,0,7/2)   0
                (1,1,1/2)   0
                (1,1,3/2)   1.414f
                (1,1,5/2)   0
                (1,1,7/2)   -1.414f
```

**Figure 8** ISOTROPY can be used to display the effect of an irrep on the occupancy of specific Wyckoff sites within either the conventional cell or indeed any unconventional cell desired. Here v bas 2,-2,0 2,2,0 -1,1,2 specifies the unit cell of the distorted structure and the command v cell 2,0,0 0,2,0 0,0,4 ensures that the effect of the irrep on the occupancy is listed for all the specified Wyckoff

sites within a representative hexagonal column, i.e. all the sites directly above the points within the basal plane of the parent structure. This makes it more intuitive to track the vacancy ordering scheme through the layers.

By doing this for all the irreps listed in Figure 7 we can determine that U1, U4, L1, $\Gamma_3^+$ and $\Gamma_1^+$ can all change occupancy on the relevant iron site, and setting OPDs as appropriate we can find their effects in the same manner as seen in Fig. 8. Using all these results we have shown that it is indeed possible to reproduce the ordering scheme exactly and with what ratio of distortions. For the 4C structure we add the irreps with mode amplitudes in the following ratios: $\frac{U1-U4}{4\sqrt{2}} + \frac{L1}{8} + \frac{\Gamma_3^+}{8} + \frac{7\Gamma_1^+}{8}$. It is noticeable that the contributions of U1 and U4 must be opposite and equal.

**Appendix B. Landau expansions for strain and order parameter coupling**

The INVARIANTS tool, part of the ISOTROPY suite, gives symmetry allowed terms in the Landau free energy expansion for a phase transition. The orthogonal axes chosen follow the convention of Nye: x // a ([100] direction), z // c ([001) direction] and y perpendicular to both ([120] direction). The expansion is given in terms of a magnetic order parameter, $Q_{Vac}$, a magnetic order parameter, $Q_{Mag}$, the six components, $e_i$, of the spontaneous strain tensor, and elastic constants, $C_{ik}^o$, of the parent hexagonal structure:

$$G = G(Q_{Mag}) + G(Q_{Vac}) + G(Q_{Mag}, Q_{Vac}) + G(Q_{Mag}, e) + G(Q_{Vac}, e) + G(Q_{Mag}, Q_{Vac}, e) + \frac{1}{2} \sum_{i,k} C_{ik}^0 e_i e_k$$

where

$$G(Q_{Mag}) = \frac{1}{2} a(T - T_{CM}) Q_{Mag}^2 + \frac{1}{4} b Q_{Mag}^4 + \frac{1}{6} c Q_{Mag}^6$$

$$G(Q_{Vac}) = \frac{1}{2} u(T - T_{CV}) Q_{Vac}^2 + \frac{1}{4} v Q_{Vac}^4 + \frac{1}{6} w Q_{Vac}^6$$

The lowest order coupling term for the two order parameters is biquadratic:

$$G(Q_{Mag}, Q_{Vac}) = \lambda_{MV} Q_{Mag}^2 Q_{Vac}^2$$

The lowest order allowed coupling terms for strain and the magnetic order parameter are:

$$\lambda_{eM1}(e_1 + e_2)Q_{Mag}^2 + \lambda_{eM2} e_3 Q_{Mag}^2 + \lambda_{eM3}\left(\frac{2}{3}(e_1 - e_2) - \frac{4}{\sqrt{3}} e_6\right)Q_{Mag}^2$$

$$\lambda_{eM4}(e_4^2 + e_5^2)Q_{Mag}^2 + \lambda_{eM5}(2e_4^2 + \frac{10}{3}e_5^2 - \frac{4}{\sqrt{3}}e_4 e_5)Q_{Mag}^2$$

The lowest order terms for coupling between strain and the order parameter for vacancy ordering are

$$\lambda_{eV1}(e_1 + e_2)Q_{Vac}^2 + \lambda_{eV2}e_3 Q_{Vac}^2 + \lambda_{eV3}\left(\frac{1}{\sqrt{3}}(e_1 - e_2) - 2e_6\right)Q_{Mag}^2$$

$$\lambda_{eV4}(e_4^2 + e_5^2)Q_{Vac}^2 + \lambda_{eV5}\left(\left(1 - \frac{\sqrt{2}}{12}\right)e_4^2 + e_5^2 - \frac{2}{4\sqrt{3}}e_4 e_5\right)Q_{Vac}^2$$

The lowest order allowed terms for coupling of strain with both order parameters are

$$\lambda_{eVM1}(e_1 + e_2)Q_{Vac}^2 Q_{Mag}^2 + \lambda_{eVM2}e_3 Q_{Vac}^2 Q_{Mag}^2 + \lambda_{eMV3}\left(\frac{2}{3}(e_1 - e_2) - \frac{4}{\sqrt{3}}e_6\right)Q_{Vac}^2 Q_{Mag}^2$$

and triquadratic for the remaining two:

$$\lambda_{eVM4}\left(\left(7 + \frac{4}{\sqrt{3}}\right)e_4^2 + 5e_5^2 + 10e_4 e_5\right)Q_{Vac}^2 Q_{Mag}^2$$

Taking these coupling terms and adding the standard Landau expansion terms for the two independent order parameters as well as the elastic energy contribution the final expression becomes

$$\begin{aligned}
G = {}& \frac{1}{2}a(T - T_C)Q_{Mag}^2 + \frac{1}{4}bQ_{Mag}^4 + \frac{1}{6}cQ_{Mag}^6 + \frac{1}{2}u(T - T_C)Q_{Vac}^2 + \frac{1}{4}vQ_{Vac}^4 + \frac{1}{6}wQ_{Vac}^6 \\
& + \lambda_{MV}Q_{Mag}^2 Q_{Vac}^2 + \lambda_{eM1}(e_1 + e_2)Q_{Mag}^2 + \lambda_{eM2}e_3 Q_{Mag}^2 \\
& + \lambda_{eM3}\left(\frac{2}{3}(e_1 - e_2) - \frac{4}{\sqrt{3}}e_6\right)Q_{Mag}^2 + \lambda_{eM4}(e_4^2 + e_5^2)Q_{Mag}^2 \\
& + \lambda_{eM5}(2e_4^2 + \frac{10}{3}e_5^2 - \frac{4}{\sqrt{3}}e_4 e_5)Q_{Mag}^2 + \lambda_{eV1}(e_1 + e_2)Q_{Vac}^2 + \lambda_{eV2}e_3 Q_{Vac}^2 \\
& + \lambda_{eV3}\left(\frac{1}{\sqrt{3}}(e_1 - e_2) - 2e_6\right)Q_{Mag}^2 + \lambda_{eV4}(e_4^2 + e_5^2)Q_{Vac}^2 \\
& + \lambda_{eV5}\left(\left(1 - \frac{\sqrt{2}}{12}\right)e_4^2 + e_5^2 - \frac{2}{4\sqrt{3}}e_4 e_5\right)Q_{Vac}^2 + \lambda_{eVM1}(e_1 + e_2)Q_{Vac}^2 Q_{Mag}^2 \\
& + \lambda_{eVM2}e_3 Q_{Vac}^2 Q_{Mag}^2 + \lambda_{eMV3}\left(\frac{2}{3}(e_1 - e_2) - \frac{4}{\sqrt{3}}e_6\right)Q_{Vac}^2 Q_{Mag}^2 \\
& + \lambda_{eVM4}\left(\left(7 + \frac{4}{\sqrt{3}}\right)e_4^2 + 5e_5^2 + 10e_4 e_5\right)Q_{Vac}^2 Q_{Mag}^2 + \frac{1}{2}\sum_{i,k}C_{ik}^0 e_i e_k
\end{aligned}$$

**S1. Tables in full**

**S1.1. 4C**

**S1.1.1. 4C U1 (1/2,0,1/4)**

**Table S1** Order parameter components and unit cell configurations for subgroups of $P6_3/mmc$ which can arise from phase transitions in which irrep U1(1/2,0,1/4) is the active representation.

| Space Group | Space group Number | OPD Name | OPD vector | Basis Vectors | Origin |
|---|---|---|---|---|---|
| C2/m | 12 | P1 | (a,0,0,0,0,0) | (2,1,-4),(0,1,0),(2,1,0) | (0,0,0) |
| Imm2 | 44 | P2 | (a,-0.414a,0,0,0,0) | (0,0,4),(0,1,0),(-2,-1,0) | (-7/8,1/16,-7/4) |
| C2/m | 12 | P3 | (a,0,a,0,0,0) | (2,-2,0),(2,2,0),(-1,1,2) | (0,0,0) |
| C2/c | 15 | P4 | (a,0,0,a,0,0) | (2,-2,0),(2,2,0),(-1,1,2) | (0,1/2,0) |
| Fmm2 | 42 | P5 | (a,-0.414a,a,-0.414a,0,0) | (0,0,4),(2,2,0),(-2,2,0) | (-1/8,1/8,1/4) |
| Fdd2 | 43 | P6 | (a,-0.414a,0.414a,a,0,0) | (0,0,4),(2,2,0),(-2,2,0) | (-7/8,7/8,7/4) |
| P$\bar{3}$m1 | 164 | P7 | (a,0,a,0,a,0) | (2,0,0),(0,2,0),(0,0,4) | (0,0,0) |
| P$\bar{6}$m2 | 187 | P8 | (a,-0.414a,a,-0.414a,a,-0.414a) | (0,-2,0),(2,2,0),(0,0,4) | (0,0,1/4) |
| Cm | 8 | C1 | (a,b,0,0,0,0) | (2,1,-4),(0,1,0),(2,1,0) | (0,0,0) |
| C2 | 5 | C2 | (a,b,a,-b,0,0) | (2,-2,0),(2,2,0),(-1,1,2) | (0,0,0) |
| C2 | 5 | C3 | (a,b,0.707a0.707b,-0.707a0.707b,0,0) | (2,2,0),(-2,2,0),(1,1,2) | (1/8,15/8,1/4) |
| P$\bar{1}$ | 2 | C4 | (a,0,b,0,0,0) | (1,1,2),(0,2,0),(-2,0,0) | (0,0,0) |
| P$\bar{1}$ | 2 | C5 | (a,0,0,b,0,0) | (1,1,2),(0,2,0),(-2,0,0) | (0,1/2,0) |
| Cm | 8 | C6 | (a,-0.414a,b,-0.414b,0,0) | (-2,-2,0),(0,0,4),(-2,0,0) | (-1/8,-1/8,1/4) |
| Cc | 9 | C7 | (a,-0.414a,b,2.414b,0,0) | (-2,-2,0),(0,0,4),(-2,0,0) | (-5/8,-5/8,5/4) |
| Cm | 8 | C8 | (a,b,a,b,0,0) | (2,-2,0),(2,2,0),(-1,1,2) | (0,0,0) |
| Cc | 9 | C9 | (a,b,b,-a,0,0) | (2,-2,0),(2,2,0),(-1,1,2) | (3/2,0,0) |
| P3m1 | 156 | C10 | (a,b,a,b,a,b) | (2,0,0),(0,2,0),(0,0,4) | (0,0,0) |
| C2/m | 12 | C11 | (a,0,b,0,a,0) | (-2,-4,0),(2,0,0),(0,0,4) | (0,0,0) |
| C2/m | 12 | C12 | (0,a,b,0,0,-a) | (-2,-4,0),(2,0,0),(0,0,4) | (1/2,0,0) |
| Amm2 | 38 | C13 | (a,-0.414a,b,-0.414b,b,-0.414b) | (0,0,4),(0,2,0),(-4,-2,0) | (0,0,1/4) |
| Abm2 | 39 | C14 | (a,-0.414a,b,2.414b,-b,2.414b) | (0,0,4),(0,2,0),(-4,-2,0) | (0,0,1/4) |
| C2 | 5 | S1 | (a,b,c,0,a,-b) | (-2,-4,0),(2,0,0),(0,0,4) | (0,0,0) |
| C2 | 5 | S2 | (a,-0.414a,b,c,0.707b0.707c,-0.707b0.707c) | (0,-2,0),(4,2,0),(0,0,4) | (0,0,1/4) |
| P$\bar{1}$ | 2 | S3 | (a,0,b,0,c,0) | (0,0,4),(0,2,0),(-2,0,0) | (0,0,0) |
| P$\bar{1}$ | 2 | S4 | (a,0,0,b,0,c) | (0,0,4),(0,2,0),(-2,0,0) | (0,1/2,0) |
| Pm | 6 | S5 | (a,-0.414a,b,-0.414b,c,-0.414c) | (-2,0,0),(0,0,4),(0,2,0) | (0,0,1/4) |
| Pc | 7 | S6 | (a,-0.414a,b,2.414b,c,2.414c) | (-2,0,0),(0,0,4),(0,2,0) | (0,0,1/4) |
| P1 | 1 | 4D1 | (a,b,c,d,0,0) | (1,1,2),(0,2,0),(-2,0,0) | (0,0,0) |
| Cm | 8 | 4D2 | (a,b,c,d,a,b) | (-2,-4,0),(2,0,0),(0,0,4) | (0,0,0) |
| P1 | 1 | 6D1 | (a,b,c,d,e,f) | (0,0,4),(0,2,0),(-2,0,0) | (0,0,0) |

### S1.1.2. 4C mGm4+ with U1

**Table S2**  Magnetic subgroups arising from coupling of irreps m$\Gamma_4^+$ and U1(1/2,0,1/4), with respect to the parent space group P6$_3$/mmc.

| SGN.M | Space group | OPD | PD Vector | Basis Vectors | Origin |
|---|---|---|---|---|---|
| 12.62 | C2'/m' | P1(1)P1(1) | (a,0,0,0,0,b) | (2,1,-4),(0,1,0),(2,1,0) | (0,0,0) |
| 44.232 | Im'm'2 | P2(1)P1(1) | (a,-0.414a,0,0,0,b) | (0,0,4),(0,1,0),(-2,-1,0) | (-7/8,1/16,-7/4) |
| 12.62 | C2'/m' | P3(1)P1(1) | (a,0,a,0,0,b) | (2,-2,0),(2,2,0),(-1,1,2) | (0,0,0) |
| 15.89 | C2'/c' | P4(1)P1(1) | (a,0,0,a,0,b) | (2,-2,0),(2,2,0),(-1,1,2) | (0,1/2,0) |
| 42.222 | Fm'm'2 | P5(1)P1(1) | (a,-0.414a,a,-0.414a,0,b) | (0,0,4),(2,2,0),(-2,2,0) | (-1/8,1/8,1/4) |
| 43.227 | Fd'd'2 | P6(1)P1(1) | (a,-0.414a,0.414a,a,0,b) | (0,0,4),(2,2,0),(-2,2,0) | (-7/8,7/8,7/4) |
| 164.89 | P$\bar{3}$m'1 | P7(1)P1(1) | (a,0,a,0,a,0,b) | (0,-2,0),(2,2,0),(0,0,4) | (0,0,0) |
| 187.211 | P$\bar{6}$'m'2 | P8(1)P1(1) | (a,-0.414a,a,-0.414a,a,-0.414a,b) | (2,2,0),(-2,0,0),(0,0,4) | (0,0,1/4) |
| 8.34 | Cm' | C1(1)P1(1) | (a,b,0,0,0,0,c) | (2,1,-4),(0,1,0),(2,1,0) | (0,0,0) |
| 5.15 | C2' | C2(1)P1(1) | (a,b,a,-b,0,0,c) | (2,-2,0),(2,2,0),(-1,1,2) | (0,0,0) |
| 5.13 | C2 | C3(1)P1(1) | (a,b,0.707a-0.707b,-0.707a-0.707b,0,0,c) | (2,2,0),(-2,2,0),(1,1,2) | (1/8,15/8,1/4) |
| 2.4 | P$\bar{1}$ | C4(1)P1(1) | (a,0,b,0,0,0,c) | (1,1,2),(0,2,0),(-2,0,0) | (0,0,0) |
| 2.4 | P$\bar{1}$ | C5(1)P1(1) | (a,0,0,b,0,0,c) | (1,1,2),(0,2,0),(-2,0,0) | (0,1/2,0) |
| 8.34 | Cm' | C6(1)P1(1) | (a,-0.414a,b,-0.414b,0,0,c) | (-2,-2,0),(0,0,4),(-2,0,0) | (-1/8,-1/8,1/4) |
| 9.39 | Cc' | C7(1)P1(1) | (a,-0.414a,b,2.414b,0,0,c) | (-2,-2,0),(0,0,4),(-2,0,0) | (-5/8,-5/8,5/4) |
| 8.34 | Cm' | C8(1)P1(1) | (a,b,a,b,0,0,c) | (2,-2,0),(2,2,0),(-1,1,2) | (0,0,0) |
| 9.39 | Cc' | C9(1)P1(1) | (a,b,b,-a,0,0,c) | (2,-2,0),(2,2,0),(-1,1,2) | (3/2,0,0) |
| 156.51 | P3m'1 | C10(1)P1(1) | (a,b,a,b,a,b,c) | (0,-2,0),(2,2,0),(0,0,4) | (0,0,0) |
| 12.62 | C2'/m' | C11(1)P1(1) | (a,0,b,0,a,0,c) | (-2,-4,0),(2,0,0),(0,0,4) | (0,0,0) |
| 12.62 | C2'/m' | C12(1)P1(1) | (0,a,b,0,0,-a,c) | (-2,-4,0),(2,0,0),(0,0,4) | (1/2,0,0) |
| 38.191 | Am'm'2 | C13(1)P1(1) | (a,-0.414a,b,-0.414b,b,-0.414b,c) | (0,0,4),(0,2,0),(-4,-2,0) | (0,0,1/4) |
| 39.199 | Ab'm'2 | C14(1)P1(1) | (a,-0.414a,b,2.414b,-b,-2.414b,c) | (0,0,4),(0,2,0),(-4,-2,0) | (0,0,1/4) |
| 5.15 | C2' | S1(1)P1(1) | (a,b,c,0,a,-b,d) | (-2,-4,0),(2,0,0),(0,0,4) | (0,0,0) |
| 5.13 | C2 | S2(1)P1(1) | (a,-0.414a,b,c,0.707b-0.707c,-0.707b-0.707c,d) | (0,-2,0),(4,2,0),(0,0,4) | (0,0,1/4) |
| 2.4 | P$\bar{1}$ | S3(1)P1(1) | (a,0,b,0,c,0,d) | (0,0,4),(0,2,0),(-2,0,0) | (0,0,0) |
| 2.4 | P$\bar{1}$ | S4(1)P1(1) | (a,0,0,b,0,c,d) | (0,0,4),(0,2,0),(-2,0,0) | (0,1/2,0) |
| 6.2 | Pm' | S5(1)P1(1) | (a,-0.414a,b,-0.414b,c,-0.414c,d) | (-2,0,0),(0,0,4),(0,2,0) | (0,0,1/4) |
| 7.26 | Pc' | S6(1)P1(1) | (a,-0.414a,b,2.414b,c,2.414c,d) | (-2,0,0),(0,0,4),(0,2,0) | (0,0,1/4) |
| 1.1 | P1 | 4D1(1)P1(1) | (a,b,c,d,0,0,e) | (1,1,2),(0,2,0),(-2,0,0) | (0,0,0) |
| 8.34 | Cm' | 4D2(1)P1(1) | (a,b,c,d,a,b,e) | (-2,-4,0),(2,0,0),(0,0,4) | (0,0,0) |
| 1.1 | P1 | 6D1(1)P1(1) | (a,b,c,d,e,f,g) | (0,0,4),(0,2,0),(-2,0,0) | (0,0,0) |

**S1.1.3. 4C mGm5+ and U1**

**Table S3** Magnetic subgroups arising from coupling of irreps $m\Gamma_5^+$ and $U1(1/2,0,1/4)$, with respect to the parent space group $P6_3/mmc$.

| SGN.MSGN | Subgroup | OPD | OPD vector | Basis Vectors | Origin |
|---|---|---|---|---|---|
| 12.58 | C2/m | P1(1)P1(1) | (a,0,0,0,0,b,-1.732b) | (2,1,-4),(0,1,0),(2,1,0) | (0,0,0) |
| 44.229 | Imm2 | P2(1)P1(1) | (a,-0.414a,0,0,0,b,-1.732b) | (0,1,0),(0,0,4),(2,1,0) | (0,0,1/4) |
| 12.58 | C2/m | P3(1)P1(3) | (a,0,a,0,0,b,1.732b) | (2,-2,0),(2,2,0),(-1,1,2) | (0,0,0) |
| 15.85 | C2/c | P4(1)P1(3) | (a,0,0,a,0,b,1.732b) | (2,-2,0),(2,2,0),(-1,1,2) | (0,1/2,0) |
| 42.219 | Fmm2 | P5(1)P1(3) | (a,-0.414a,a,-0.414a,0,b,1.732b) | (2,2,0),(0,0,4),(2,-2,0) | (1/8,-1/8,1/4) |
| 43.224 | Fdd2 | P6(1)P1(3) | (a,-0.414a,0.414a,a,0,b,1.732b) | (2,2,0),(0,0,4),(2,-2,0) | (11/8,-3/8,3/4) |
| 8.32 | Cm | C1(1)P1(1) | (a,b,0,0,0,0,c,-1.732c) | (2,1,-4),(0,1,0),(2,1,0) | (0,0,0) |
| 5.13 | C2 | C2(1)P1(3) | (a,b,a,-b,0,0,c,1.732c) | (2,-2,0),(2,2,0),(-1,1,2) | (0,0,0) |
| 5.13 | C2 | C3(1)P1(3) | (a,b,0.707a-0.707b,-0.707a-0.707b,0,0,c,1.732c) | (2,2,0),(-2,2,0),(1,1,2) | (1/8,15/8,1/4) |
| 8.32 | Cm | C8(1)P1(3) | (a,b,a,b,0,0,c,1.732c) | (2,-2,0),(2,2,0),(-1,1,2) | (0,0,0) |
| 9.37 | Cc | C9(1)P1(3) | (a,b,b,-a,0,0,c,1.732c) | (2,-2,0),(2,2,0),(-1,1,2) | (3/2,0,0) |
| 12.58 | C2/m | C11(1)P1(2) | (a,0,b,0,a,0,-2c,0) | (-2,-4,0),(2,0,0),(0,0,4) | (0,0,0) |
| 12.58 | C2/m | C12(1)P1(2) | (0,a,b,0,0,-a,-2c,0) | (-2,-4,0),(2,0,0),(0,0,4) | (1/2,0,0) |
| 38.187 | Amm2 | C13(1)P1(1) | (a,-0.414a,b,-0.414b,b,-0.414b,c,-1.732c) | (0,0,4),(0,2,0),(-4,-2,0) | (0,0,1/4) |
| 39.195 | Abm2 | C14(1)P1(1) | (a,-0.414a,b,2.414b,-b,-2.414b,c,-1.732c) | (0,0,4),(0,2,0),(-4,-2,0) | (0,0,1/4) |
| 5.13 | C2 | S1(1)P1(2) | (a,b,c,0,a,-b,-2d,0) | (-2,-4,0),(2,0,0),(0,0,4) | (0,0,0) |
| 5.13 | C2 | S2(1)P1(1) | (a,-0.414a,b,c,0.707b-0.707c,-0.707b-0.707c,d,-1.732d) | (0,-2,0),(4,2,0),(0,0,4) | (0,0,1/4) |
| 8.32 | Cm | 4D2(1)P1(2) | (a,b,c,d,a,b,-2e,0) | (-2,-4,0),(2,0,0),(0,0,4) | (0,0,0) |
| 12.62 | C2'/m' | P1(1)P2(1) | (a,0,0,0,0,0,b,0.577b) | (2,1,-4),(0,1,0),(2,1,0) | (0,0,0) |
| 44.231 | Im'm2' | P2(1)P2(1) | (a,-0.414a,0,0,0,0,b,0.577b) | (0,1,0),(0,0,4),(2,1,0) | (0,0,1/4) |
| 12.62 | C2'/m' | P3(1)P2(3) | (a,0,a,0,0,0,-b,0.577b) | (2,-2,0),(2,2,0),(-1,1,2) | (0,0,0) |
| 15.89 | C2'/c' | P4(1)P2(3) | (a,0,0,a,0,0,-b,0.577b) | (2,-2,0),(2,2,0),(-1,1,2) | (0,1/2,0) |
| 42.221 | Fm'm2' | P5(1)P2(3) | (a,-0.414a,a,-0.414a,0,0,-b,0.577b) | (2,2,0),(0,0,4),(2,-2,0) | (1/8,-1/8,1/4) |
| 43.226 | Fd'd2' | P6(1)P2(3) | (a,-0.414a,0.414a,a,0,0,-b,0.577b) | (2,2,0),(0,0,4),(2,-2,0) | (11/8,-3/8,3/4) |
| 8.34 | Cm' | C1(1)P2(1) | (a,b,0,0,0,0,c,0.577c) | (2,1,-4),(0,1,0),(2,1,0) | (0,0,0) |
| 5.15 | C2' | C2(1)P2(3) | (a,b,a,-b,0,0,-c,0.577c) | (2,-2,0),(2,2,0),(-1,1,2) | (0,0,0) |
| 5.15 | C2' | C3(1)P2(3) | (a,b,0.707a-0.707b,-0.707a-0.707b,0,0,-c,0.577c) | (2,2,0),(-2,2,0),(1,1,2) | (1/8,15/8,1/4) |
| 8.34 | Cm' | C8(1)P2(3) | (a,b,a,b,0,0,-c,0.577c) | (2,-2,0),(2,2,0),(-1,1,2) | (0,0,0) |
| 9.39 | Cc' | C9(1)P2(3) | (a,b,b,-a,0,0,-c,0.577c) | (2,-2,0),(2,2,0),(-1,1,2) | (3/2,0,0) |
| 12.62 | C2'/m' | C11(1)P2(2) | (a,0,b,0,a,0,0,-1.155c) | (-2,-4,0),(2,0,0),(0,0,4) | (0,0,0) |
| 12.62 | C2'/m' | C12(1)P2(2) | (0,a,b,0,0,-a,0,-1.155c) | (-2,-4,0),(2,0,0),(0,0,4) | (1/2,0,0) |
| 38.19 | Amm'2' | C13(1)P2(1) | (a,-0.414a,b,-0.414b,b,- | (0,0,4),(0,2,0),(-4,-2,0) | (0,0,1/4) |
| | | | 0.414b,c,0.577c) | | |
| 39.198 | Abm'2' | C14(1)P2(1) | (a,-0.414a,b,2.414b,-b,-2.414b,c,0.577c) | (0,0,4),(0,2,0),(-4,-2,0) | (0,0,1/4) |
| 5.15 | C2' | S1(1)P2(2) | (a,b,c,0,a,-b,0,-1.155d) | (-2,-4,0),(2,0,0),(0,0,4) | (0,0,0) |
| 5.15 | C2' | S2(1)P2(1) | (a,-0.414a,b,c,0.707b-0.707c,-0.707b-0.707c,d,0.577d) | (0,-2,0),(4,2,0),(0,0,4) | (0,0,1/4) |
| 8.34 | Cm' | 4D2(1)P2(2) | (a,b,c,d,a,b,0,-1.155e) | (-2,-4,0),(2,0,0),(0,0,4) | (0,0,0) |
| 2.4 | P$\bar{1}$ | P1(1)C1(1) | (a,0,0,0,0,0,b,c) | (1,0,2),(0,1,0),(-2,0,0) | (0,0,0) |
| 8.32 | Cm | P2(1)C1(1) | (a,-0.414a,0,0,0,0,b,c) | (-2,0,0),(0,0,4),(0,1,0) | (-1/8,0,1/4) |
| 1.1 | P1 | C1(1)C1(1) | (a,b,0,0,0,0,c,d) | (1,0,2),(0,1,0),(-2,0,0) | (0,0,0) |
| 2.4 | P$\bar{1}$ | C4(1)C1(1) | (a,0,b,0,0,0,c,d) | (1,1,2),(0,2,0),(-2,0,0) | (0,0,0) |
| 2.4 | P$\bar{1}$ | C5(1)C1(1) | (a,0,0,b,0,0,c,d) | (1,1,2),(0,2,0),(-2,0,0) | (0,1/2,0) |
| 8.32 | Cm | C6(1)C1(1) | (a,-0.414a,b,-0.414b,0,0,c,d) | (-2,-2,0),(0,0,4),(-2,0,0) | (-1/8,-1/8,1/4) |
| 9.37 | Cc | C7(1)C1(1) | (a,-0.414a,b,2.414b,0,0,c,d) | (-2,-2,0),(0,0,4),(-2,0,0) | (-5/8,-5/8,5/4) |
| 2.4 | P$\bar{1}$ | S3(1)C1(1) | (a,0,b,0,c,0,d,e) | (0,0,4),(0,2,0),(-2,0,0) | (0,0,0) |
| 2.4 | P$\bar{1}$ | S4(1)C1(1) | (a,0,0,b,0,c,d,e) | (0,0,4),(0,2,0),(-2,0,0) | (0,1/2,0) |
| 6.18 | Pm | S5(1)C1(1) | (a,-0.414a,b,-0.414b,c,-0.414c,d,e) | (-2,0,0),(0,0,4),(0,2,0) | (0,0,1/4) |
| 7.24 | Pc | S6(1)C1(1) | (a,-0.414a,b,2.414b,c,2.414c,d,e) | (-2,0,0),(0,0,4),(0,2,0) | (0,0,1/4) |
| 1.1 | P1 | 4D1(1)C1(1) | (a,b,c,d,0,0,e,f) | (1,1,2),(0,2,0),(-2,0,0) | (0,0,0) |
| 1.1 | P1 | 6D1(1)C1(1) | (a,b,c,d,e,f,g,h) | (0,0,4),(0,2,0),(-2,0,0) | (0,0,0) |

## S2. 5C

### S2.1. U1

**Table S4** Order parameter components and unit cell configurations for subgroups of $P6_3/mmc$ which can arise from phase transitions in which irrep U1(1/2,0,1/5) is the active representation.

| Space Group | Space group Number | OPD Name | OPD vector | Basis Vectors | Origin |
|---|---|---|---|---|---|
| 58 | Pnnm | P1 | (a,0,0,0,0,0) | (0,0,5),(2,1,0),(0,1,0) | (0,0,0) |
| 59 | Pmmn | P2 | (0,a,0,0,0,0) | (0,0,5),(0,-1,0),(2,1,0) | (1/2,1/2,0) |
| 164 | P$\bar{3}$m1 | P3 | (a,0,a,0,a,0) | (2,0,0),(0,2,0),(0,0,5) | (0,0,0) |
| 187 | P$\bar{6}$m2 | P4 | (a,-0.325a,a,-0.325a,a,-0.325a) | (0,-2,0),(2,2,0),(0,0,5) | (0,0,1/4) |
| 64 | Cmca | P5 | (a,0,0,0,a,0) | (2,0,0),(2,4,0),(0,0,5) | (0,0,0) |
| 63 | Cmcm | P6 | (0,a,0,0,0,-a) | (2,0,0),(2,4,0),(0,0,5) | (1/2,0,0) |
| 31 | Pmn2$_1$ | C1 | (a,b,0,0,0,0) | (0,-1,0),(2,1,0),(0,0,5) | (1/2,1/4,0) |
| 156 | P3m1 | C2 | (a,b,a,b,a,b) | (2,0,0),(0,2,0),(0,0,5) | (0,0,0) |
| 20 | C222$_1$ | C3 | (a,b,0,0,a,-b) | (2,0,0),(2,4,0),(0,0,5) | (1/2,0,0) |
| 14 | P2$_1$/c | C4 | (0,0,a,0,b,0) | (-2,0,0),(0,0,5),(0,2,0) | (0,0,0) |
| 11 | P2$_1$/m | C5 | (0,0,0,a,0,b) | (-2,0,0),(0,0,5),(0,2,0) | (0,1/2,0) |
| 14 | P2$_1$/c | C6 | (a,0,0,0,0,b) | (-2,0,0),(0,0,5),(2,2,0) | (0,1/2,0) |
| 12 | C2/m | C7 | (a,0,b,0,a,0) | (-2,-4,0),(2,0,0),(0,0,5) | (0,0,0) |
| 12 | C2/m | C8 | (0,a,b,0,0,-a) | (-2,-4,0),(2,0,0),(0,0,5) | (1/2,0,0) |
| 38 | Amm2 | C9 | (a,-0.325a,b,-0.325b,b,-0.325b) | (0,0,5),(0,2,0),(-4,-2,0) | (0,0,1/4) |

| SGN | Space group | OPD Name | OPD vector | Basis Vectors | Origin |
|---|---|---|---|---|---|
| 39 | Abm2 | C10 | (a,-0.325a,b,3.078b,-b,-3.078b) | (0,0,5),(0,2,0),(-4,-2,0) | (0,0,1/4) |
| 36 | Cmc2$_1$ | C11 | (a,b,0,0,a,b) | (2,0,0),(2,4,0),(0,0,5) | (1/2,-1,0) |
| 5 | C2 | S1 | (a,b,c,0,a,-b) | (-2,-4,0),(2,0,0),(0,0,5) | (0,0,0) |
| 5 | C2 | S2 | (a,-0.325a,b,c,0.809b-0.588c,-0.588b-0.809c) | (0,-2,0),(4,2,0),(0,0,5) | (0,0,1/4) |
| 2 | P$\bar{1}$ | S3 | (a,0,b,0,c,0) | (0,0,5),(0,2,0),(-2,0,0) | (0,0,0) |
| 2 | P$\bar{1}$ | S4 | (a,0,0,b,0,c) | (0,0,5),(0,2,0),(-2,0,0) | (0,1/2,0) |
| 6 | Pm | S5 | (a,-0.325a,b,-0.325b,c,-0.325c) | (-2,0,0),(0,0,5),(0,2,0) | (0,0,1/4) |
| 7 | Pc | S6 | (a,-0.325a,b,3.078b,c,3.078c) | (-2,0,0),(0,0,5),(0,2,0) | (0,0,1/4) |
| 4 | P2$_1$ | 4D1 | (0,0,a,b,c,d) | (-2,0,0),(0,0,5),(0,2,0) | (0,1/2,0) |
| 8 | Cm | 4D2 | (a,b,c,d,a,b) | (-2,-4,0),(2,0,0),(0,0,5) | (0,0,0) |
| 1 | P1 | 6D1 | (a,b,c,d,e,f) | (0,0,5),(0,2,0),(-2,0,0) | (0,0,0) |
| 58 | Pnnm | P1 | (a,0,0,0,0,0) | (0,0,5),(2,1,0),(0,1,0) | (0,0,0) |
| 59 | Pmmn | P2 | (0,a,0,0,0,0) | (0,0,5),(0,-1,0),(2,1,0) | (1/2,1/2,0) |
| 164 | P$\bar{3}$m1 | P3 | (a,0,a,0,a,0) | (2,0,0),(0,2,0),(0,0,5) | (0,0,0) |

## S2.2. U1 and mGm4+

**Table S5** Magnetic subgroups arising from coupling of irreps m$\Gamma_4^+$ and U1(1/2,0,1/5), with respect to the parent space group P6$_3$/mmc.

| SGN.m | Space group | OPD Name | OPD vector | Basis Vectors | Origin |
|---|---|---|---|---|---|
| 58.398 | Pnn'm' | P1(1)P1(1) | (a,0,0,0,0,0,b) | (-2,-1,0),(0,0,5),(0,1,0) | (0,0,0) |
| 59.409 | Pm'm'n | P2(1)P1(1) | (0,a,0,0,0,0,b) | (0,0,5),(0,-1,0),(2,1,0) | (1/2,1/2,0) |
| 164.89 | P$\bar{3}$m'1 | P3(1)P1(1) | (a,0,a,0,a,0,b) | (0,-2,0),(2,2,0),(0,0,5) | (0,0,0) |
| 187.211 | P$\bar{6}$'m'2 | P4(1)P1(1) | (a,-0.325a,a,-0.325a,a,-0.325a,b) | (2,2,0),(-2,0,0),(0,0,5) | (0,0,1/4) |
| 64.476 | Cm'ca' | P5(1)P1(1) | (a,0,0,0,a,0,b) | (2,0,0),(2,4,0),(0,0,5) | (0,0,0) |
| 63.464 | Cm'cm' | P6(1)P1(1) | (0,a,0,0,0,-a,b) | (2,0,0),(2,4,0),(0,0,5) | (1/2,0,0) |
| 31.125 | Pm'n2$_1$' | C1(1)P1(1) | (a,b,0,0,0,0,c) | (0,-1,0),(2,1,0),(0,0,5) | (1/2,1/4,0) |
| 156.51 | P3m'1 | C2(1)P1(1) | (a,b,a,b,a,b,c) | (0,-2,0),(2,2,0),(0,0,5) | (0,0,0) |
| 20.34 | C22'2$_1$' | C3(1)P1(1) | (a,b,0,0,a,-b,c) | (2,4,0),(-2,0,0),(0,0,5) | (3/2,2,5/4) |
| 14.79 | P2$_1$'/c' | C4(1)P1(1) | (0,0,a,0,b,0,c) | (-2,0,0),(0,0,5),(0,2,0) | (0,0,0) |
| 11.54 | P2$_1$'/m' | C5(1)P1(1) | (0,0,0,a,0,b,c) | (-2,0,0),(0,0,5),(0,2,0) | (0,1/2,0) |
| 14.79 | P2$_1$'/c' | C6(1)P1(1) | (a,0,0,0,0,b,c) | (-2,0,0),(0,0,5),(2,2,0) | (0,1/2,0) |
| 12.62 | C2'/m' | C7(1)P1(1) | (a,0,b,0,a,0,c) | (-2,-4,0),(2,0,0),(0,0,5) | (0,0,0) |
| 12.62 | C2'/m' | C8(1)P1(1) | (0,a,b,0,0,-a,c) | (-2,-4,0),(2,0,0),(0,0,5) | (1/2,0,0) |
| 38.191 | Am'm'2 | C9(1)P1(1) | (a,-0.325a,b,-0.325b,b,-0.325b,c) | (0,0,5),(0,2,0),(-4,-2,0) | (0,0,1/4) |
| 39.199 | Ab'm'2 | C10(1)P1(1) | (a,-0.325a,b,3.078b,-b,-3.078b,c) | (0,0,5),(0,2,0),(-4,-2,0) | (0,0,1/4) |
| 36.174 | Cm'c2$_1$' | C11(1)P1(1) | (a,b,0,0,a,b,c) | (2,0,0),(2,4,0),(0,0,5) | (1/2,-1,0) |
| 5.15 | C2' | S1(1)P1(1) | (a,b,c,0,a,-b,d) | (-2,-4,0),(2,0,0),(0,0,5) | (0,0,0) |
| 5.13 | C2 | S2(1)P1(1) | (a,-0.325a,b,c,0.809b-0.588c,-0.588b-0.809c,d) | (0,-2,0),(4,2,0),(0,0,5) | (0,0,1/4) |
| 2.4 | P$\bar{1}$ | S3(1)P1(1) | (a,0,b,0,c,0,d) | (0,0,5),(0,2,0),(-2,0,0) | (0,0,0) |

| SGN.m | Space group | OPD Name | OPD vector | Basis Vectors | Origin |
|---|---|---|---|---|---|
| 2.4 | P$\bar{1}$ | S4(1)P1(1) | (a,0,0,b,0,c,d) | (0,0,5),(0,2,0),(-2,0,0) | (0,1/2,0) |
| 6.2 | Pm' | S5(1)P1(1) | (a,-0.325a,b,-0.325b,c,-0.325c,d) | (-2,0,0),(0,0,5),(0,2,0) | (0,0,1/4) |
| 7.26 | Pc' | S6(1)P1(1) | (a,-0.325a,b,3.078b,c,3.078c,d) | (-2,0,0),(0,0,5),(0,2,0) | (0,0,1/4) |
| 4.9 | P2$_1$' | 4D1(1)P1(1) | (0,0,a,b,c,d,e) | (-2,0,0),(0,0,5),(0,2,0) | (0,1/2,0) |
| 8.34 | Cm' | 4D2(1)P1(1) | (a,b,c,d,a,b,e) | (-2,-4,0),(2,0,0),(0,0,5) | (0,0,0) |
| 1.1 | P1 | 6D1(1)P1(1) | (a,b,c,d,e,f,g) | (0,0,5),(0,2,0),(-2,0,0) | (0,0,0) |

### S2.3. U1 and mGm5+

**Table S6** Magnetic subgroups arising from coupling of irreps m$\Gamma_5^+$ and U1(1/2,0,1/5), with respect to the parent space group P6$_3$/mmc.

| SGN.m | Space group | OPD Name | OPD vector | Basis Vectors | Origin |
|---|---|---|---|---|---|
| 58.393 | Pnnm | P1(1)P1(1) | (a,0,0,0,0,b,-1.732b) | (-2,-1,0),(0,0,5),(0,1,0) | (0,0,0) |
| 59.405 | Pmmn | P2(1)P1(1) | (0,a,0,0,0,b,-1.732b) | (0,1,0),(0,0,5),(2,1,0) | (1/2,0,0) |
| 64.469 | Cmca | P5(1)P1(2) | (a,0,0,0,a,0,-2b,0) | (2,0,0),(2,4,0),(0,0,5) | (0,0,0) |
| 63.457 | Cmcm | P6(1)P1(2) | (0,a,0,0,0,-a,-2b,0) | (2,0,0),(2,4,0),(0,0,5) | (1/2,0,0) |
| 31.123 | Pmn2$_1$ | C1(1)P1(1) | (a,b,0,0,0,0,c,-1.732c) | (0,-1,0),(2,1,0),(0,0,5) | (1/2,1/4,0) |
| 20.31 | C222$_1$ | C3(1)P1(2) | (a,b,0,0,a,-b,-2c,0) | (2,0,0),(2,4,0),(0,0,5) | (1/2,0,0) |
| 12.58 | C2/m | C7(1)P1(2) | (a,0,b,0,a,0,-2c,0) | (-2,-4,0),(2,0,0),(0,0,5) | (0,0,0) |
| 12.58 | C2/m | C8(1)P1(2) | (0,a,b,0,0,-a,-2c,0) | (-2,-4,0),(2,0,0),(0,0,5) | (1/2,0,0) |
| 38.187 | Amm2 | C9(1)P1(1) | (a,-0.325a,b,-0.325b,b,-0.325b,c,-1.732c) | (0,0,5),(0,2,0),(-4,-2,0) | (0,0,1/4) |
| 39.195 | Abm2 | C10(1)P1(1) | (a,-0.325a,b,3.078b,-b,-3.078b,c,-1.732c) | (0,0,5),(0,2,0),(-4,-2,0) | (0,0,1/4) |
| 36.172 | Cmc2$_1$ | C11(1)P1(2) | (a,b,0,0,a,b,-2c,0) | (2,0,0),(2,4,0),(0,0,5) | (1/2,-1,0) |
| 5.13 | C2 | S1(1)P1(2) | (a,b,c,0,a,-b,-2d,0) | (-2,-4,0),(2,0,0),(0,0,5) | (0,0,0) |
| 5.13 | C2 | S2(1)P1(1) | (a,-0.325a,b,c,0.809b-0.588c,-0.588b-0.809c,d,-1.732d) | (0,-2,0),(4,2,0),(0,0,5) | (0,0,1/4) |
| 8.32 | Cm | 4D2(1)P1(2) | (a,b,c,d,a,b,-2e,0) | (-2,-4,0),(2,0,0),(0,0,5) | (0,0,0) |
| 58.398 | Pnn'm' | P1(1)P2(1) | (a,0,0,0,0,0,b,0.577b) | (0,0,5),(2,1,0),(0,1,0) | (0,0,0) |
| 59.41 | Pmm'n' | P2(1)P2(1) | (0,a,0,0,0,0,b,0.577b) | (0,0,5),(0,-1,0),(2,1,0) | (1/2,1/2,0) |
| 64.474 | Cm'c'a | P5(1)P2(2) | (a,0,0,0,a,0,0,-1.155b) | (2,0,0),(2,4,0),(0,0,5) | (0,0,0) |
| 63.462 | Cm'c'm | P6(1)P2(2) | (0,a,0,0,0,-a,0,-1.155b) | (2,0,0),(2,4,0),(0,0,5) | (1/2,0,0) |
| 31.127 | Pm'n'2$_1$ | C1(1)P2(1) | (a,b,0,0,0,0,c,0.577c) | (0,-1,0),(2,1,0),(0,0,5) | (1/2,1/4,0) |
| 20.33 | C2'2'2$_1$ | C3(1)P2(2) | (a,b,0,0,a,-b,0,-1.155c) | (2,0,0),(2,4,0),(0,0,5) | (1/2,0,0) |
| 12.62 | C2'/m' | C7(1)P2(2) | (a,0,b,0,a,0,0,-1.155c) | (-2,-4,0),(2,0,0),(0,0,5) | (0,0,0) |
| 12.62 | C2'/m' | C8(1)P2(2) | (0,a,b,0,0,-a,0,-1.155c) | (-2,-4,0),(2,0,0),(0,0,5) | (1/2,0,0) |
| 38.19 | Amm'2' | C9(1)P2(1) | (a,-0.325a,b,-0.325b,b,-0.325b,c,0.577c) | (0,0,5),(0,2,0),(-4,-2,0) | (0,0,1/4) |
| 39.198 | Abm'2' | C10(1)P2(1) | (a,-0.325a,b,3.078b,-b,-3.078b,c,0.577c) | (0,0,5),(0,2,0),(-4,-2,0) | (0,0,1/4) |
| 36.176 | Cm'c'2$_1$ | C11(1)P2(2) | (a,b,0,0,a,b,0,-1.155c) | (2,0,0),(2,4,0),(0,0,5) | (1/2,-1,0) |
| 5.15 | C2' | S1(1)P2(2) | (a,b,c,0,a,-b,0,-1.155d) | (-2,-4,0),(2,0,0),(0,0,5) | (0,0,0) |
| 5.15 | C2' | S2(1)P2(1) | (a,-0.325a,b,c,0.809b-0.588c,-0.588b-0.809c,d,0.577d) | (0,-2,0),(4,2,0),(0,0,5) | (0,0,1/4) |
| 8.34 | Cm' | 4D2(1)P2(2) | (a,b,c,d,a,b,0,-1.155e) | (-2,-4,0),(2,0,0),(0,0,5) | (0,0,0) |
| 14.75 | P2$_1$/c | P1(1)C1(1) | (a,0,0,0,0,0,b,c) | (0,1,0),(0,0,5),(2,0,0) | (0,0,0) |
| 11.5 | P2$_1$/m | P2(1)C1(1) | (0,a,0,0,0,0,b,c) | (-2,0,0),(0,0,5),(0,1,0) | (-1/2,0,0) |
| 4.7 | P2$_1$ | C1(1)C1(1) | (a,b,0,0,0,0,c,d) | (-2,0,0),(0,0,5),(0,1,0) | (-1/2,0,0) |
| 14.75 | P2$_1$/c | C4(1)C1(1) | (0,0,a,0,b,0,c,d) | (-2,0,0),(0,0,5),(0,2,0) | (0,0,0) |
| 11.5 | P2$_1$/m | C5(1)C1(1) | (0,0,0,a,0,b,c,d) | (-2,0,0),(0,0,5),(0,2,0) | (0,1/2,0) |
| 14.75 | P2$_1$/c | C6(1)C1(1) | (a,0,0,0,0,b,c,d) | (-2,0,0),(0,0,5),(2,2,0) | (0,1/2,0) |
| 2.4 | P$\bar{1}$ | S3(1)C1(1) | (a,0,b,0,c,0,d,e) | (0,0,5),(0,2,0),(-2,0,0) | (0,0,0) |
| 2.4 | P$\bar{1}$ | S4(1)C1(1) | (a,0,0,b,0,c,d,e) | (0,0,5),(0,2,0),(-2,0,0) | (0,1/2,0) |
| 6.18 | Pm | S5(1)C1(1) | (a,-0.325a,b,-0.325b,c,-0.325c,d,e) | (-2,0,0),(0,0,5),(0,2,0) | (0,0,1/4) |
| 7.24 | Pc | S6(1)C1(1) | (a,-0.325a,b,3.078b,c,3.078c,d,e) | (-2,0,0),(0,0,5),(0,2,0) | (0,0,1/4) |
| 4.7 | P2$_1$ | 4D1(1)C1(1) | (0,0,a,b,c,d,e,f) | (-2,0,0),(0,0,5),(0,2,0) | (0,1/2,0) |
| 1.1 | P1 | 6D1(1)C1(1) | (a,b,c,d,e,f,g,h) | (0,0,5),(0,2,0),(-2,0,0) | (0,0,0) |

## S3. 6C

### S3.1. U1

**Table S7** Order parameter components and unit cell configurations for subgroups of *P*6$_3$/*mmc* which can arise from phase transitions in which irrep U1(1/2,0,1/6) is the active representation.

| Space Group Number | Space group | OPD Name | OPD Vector | Basis Vectors | Origin |
|---|---|---|---|---|---|
| 12 | C2/m | P1 | (a,0,0,0,0,0) | (2,1,-6),(0,1,0),(2,1,0) | (0,0,0) |
| 44 | Imm2 | P2 | (a,-0.268a,0,0,0,0) | (0,0,6),(0,1,0),(-2,-1,0) | (-11/12,1/24,-11/4) |
| 12 | C2/m | P3 | (a,0,a,0,0,0) | (2,-2,0),(2,2,0),(-1,1,3) | (0,0,0) |
| 15 | C2/c | P4 | (a,0,0,a,0,0) | (2,2,0),(-2,2,0),(1,1,3) | (0,1/2,0) |
| 43 | Fdd2 | P5 | (a,-0.268a,0.268a,a,0,0) | (0,0,6),(-2,2,0),(-2,-2,0) | (-4/3,-1/3,1) |
| 42 | Fmm2 | P6 | (a,-0.268a,a,-0.268a,0,0) | (0,0,6),(2,2,0),(-2,2,0) | (-1/12,1/12,1/4) |
| 164 | P$\bar{3}$m1 | P7 | (a,0,a,0,a,0) | (2,0,0),(0,2,0),(0,0,6) | (0,0,0) |
| 187 | P$\bar{6}$m2 | P8 | (a,-0.268a,a,-0.268a,a,-0.268a) | (0,-2,0),(2,2,0),(0,0,6) | (0,0,1/4) |
| 152 | P3$_1$21 | P9 | (a,1.732a,-2a,0,a,-1.732a) | (2,0,0),(0,2,0),(0,0,6) | (0,0,1) |
| 151 | P3$_1$12 | P10 | (a,a,-1.366a,0.366a,0.366a,1.366a) | (0,-2,0),(2,2,0),(0,0,6) | (0,0,5/4) |
| 8 | Cm | C1 | (a,b,0,0,0,0) | (2,1,-6),(0,1,0),(2,1,0) | (0,0,0) |
| 5 | C2 | C2 | (a,b,a,-b,0,0) | (2,-2,0),(2,2,0),(-1,1,3) | (0,0,0) |
| 5 | C2 | C3 | (a,b,0.866a0.500b,-0.500a0.866b,0,0) | (2,2,0),(-2,2,0),(1,1,3) | (1/12,23/12,1/4) |
| 2 | P$\bar{1}$ | C4 | (a,0,b,0,0,0) | (1,1,3),(0,2,0),(-2,0,0) | (0,0,0) |
| 2 | P$\bar{1}$ | C5 | (a,0,0,b,0,0) | (1,1,3),(0,2,0),(-2,0,0) | (0,1/2,0) |
| 8 | Cm | C6 | (a,-0.268a,b,-0.268b,0,0) | (-2,-2,0),(0,0,6),(-2,0,0) | (-1/12,-1/12,1/4) |
| 9 | Cc | C7 | (a,-0.268a,b,3.732b,0,0) | (-2,-2,0),(0,0,6),(-2,0,0) | (-7/12,-7/12,7/4) |
| 8 | Cm | C8 | (a,b,a,b,0,0) | (2,-2,0),(2,2,0),(-1,1,3) | (0,0,0) |
| 9 | Cc | C9 | (a,b,b,-a,0,0) | (2,2,0),(-2,2,0),(1,1,3) | (0,3/2,0) |
| 156 | P3m1 | C10 | (a,b,a,b,a,b) | (2,0,0),(0,2,0),(0,0,6) | (0,0,0) |
| 12 | C2/m | C11 | (a,0,b,0,a,0) | (-2,-4,0),(2,0,0),(0,0,6) | (0,0,0) |
| 12 | C2/m | C12 | (0,a,b,0,0,-a) | (-2,-4,0),(2,0,0),(0,0,6) | (1/2,0,0) |
| 38 | Amm2 | C13 | (a,-0.268a,b,-0.268b,b,-0.268b) | (0,0,6),(0,2,0),(-4,-2,0) | (0,0,1/4) |
| 39 | Abm2 | C14 | (a,-0.268a,b,3.732b,-b,3.732b) | (0,0,6),(0,2,0),(-4,-2,0) | (0,0,1/4) |
| 144 | P3$_1$ | C15 | (a,b,-0.500a0.866b,0.866a-0.500b,-0.500a+0.866b,-0.866a-0.500b) | (2,2,0),(-2,0,0),(0,0,6) | (0,0,0) |
| 5 | C2 | S1 | (a,b,c,0,a,-b) | (-2,-4,0),(2,0,0),(0,0,6) | (0,0,0) |
| 5 | C2 | S2 | (a,-0.268a,b,c,0.866b0.500c,-0.500b0.866c) | (0,-2,0),(4,2,0),(0,0,6) | (0,0,1/4) |
| 2 | P$\bar{1}$ | S3 | (a,0,b,0,c,0) | (0,0,6),(0,2,0),(-2,0,0) | (0,0,0) |
| 2 | P$\bar{1}$ | S4 | (a,0,0,b,0,c) | (0,0,6),(0,2,0),(-2,0,0) | (0,1/2,0) |
| 6 | Pm | S5 | (a,-0.268a,b,-0.268b,c,-0.268c) | (-2,0,0),(0,0,6),(0,2,0) | (0,0,1/4) |
| 7 | Pc | S6 | (a,-0.268a,b,3.732b,c,3.732c) | (-2,0,0),(0,0,6),(0,2,0) | (0,0,1/4) |
| 1 | P1 | 4D1 | (a,b,c,d,0,0) | (1,1,3),(0,2,0),(-2,0,0) | (0,0,0) |
| 8 | Cm | 4D2 | (a,b,c,d,a,b) | (-2,-4,0),(2,0,0),(0,0,6) | (0,0,0) |

| | | | | | | |
|---|---|---|---|---|---|---|
| 1 | P1 | 6D1 | (a,b,c,d,e,f) | (0,0,6),(0,2,0),(-2,0,0) | (0,0,0) | |

### S3.1.1. U1 and gM4+

**Table S8** Magnetic subgroups arising from coupling of irreps $m\Gamma_4^+$ and U1(1/2,0,1/6), with respect to the parent space group P6$_3$/mmc.

| Space Group Number | Space group | OPD Name | OPD Vector | Basis Vectors | Origin |
|---|---|---|---|---|---|
| 12.62 | C2'/m' | P1(1)P1(1) | (a,0,0,0,0,b) | (2,1,-6),(0,1,0),(2,1,0) | (0,0,0) |
| 44.232 | Im'm'2 | P2(1)P1(1) | (a,-0.268a,0,0,0,b) | (0,0,6),(0,1,0),(-2,-1,0) | (-11/12,1/24,-11/4) |
| 12.62 | C2'/m' | P3(1)P1(1) | (a,0,a,0,0,b) | (2,-2,0),(2,2,0),(-1,1,3) | (0,0,0) |
| 15.85 | C2/c | P4(1)P1(1) | (a,0,0,a,0,b) | (2,2,0),(-2,2,0),(1,1,3) | (0,1/2,0) |
| 43.226 | Fd'd2' | P5(1)P1(1) | (a,-0.268a,0.268a,a,0,b) | (0,0,6),(-2,2,0),(-2,-2,0) | (-4/3,-1/3,1) |
| 42.222 | Fm'm'2 | P6(1)P1(1) | (a,-0.268a,a,-0.268a,0,b) | (0,0,6),(2,2,0),(-2,2,0) | (-1/12,1/12,1/4) |
| 164.89 | P$\bar{3}$m'1 | P7(1)P1(1) | (a,0,a,0,a,0,b) | (0,-2,0),(2,2,0),(0,0,6) | (0,0,0) |
| 187.211 | P$\bar{6}$'m'2 | P8(1)P1(1) | (a,-0.268a,a,-0.268a,a,-0.268a,b) | (2,2,0),(-2,0,0),(0,0,6) | (0,0,1/4) |
| 152.35 | P3$_1$2'1 | P9(1)P1(1) | (a,1.732a,-2a,0,a,-1.732a,b) | (0,-2,0),(2,2,0),(0,0,6) | (0,0,0) |
| 151.29 | P3$_1$12 | P10(1)P1(1) | (a,a,-1.366a,0.366a,0.366a,-1.366a,b) | (2,2,0),(-2,0,0),(0,0,6) | (0,0,1/4) |
| 8.34 | Cm' | C1(1)P1(1) | (a,b,0,0,0,0,c) | (2,1,-6),(0,1,0),(2,1,0) | (0,0,0) |
| 5.15 | C2' | C2(1)P1(1) | (a,b,a,-b,0,0,c) | (2,-2,0),(2,2,0),(-1,1,3) | (0,0,0) |
| 5.13 | C2 | C3(1)P1(1) | (a,b,0.866a-0.500b,-0.500a-0.866b,0,0,c) | (2,2,0),(-2,2,0),(1,1,3) | (1/12,23/12,1/4) |
| 2.4 | P$\bar{1}$ | C4(1)P1(1) | (a,0,b,0,0,c) | (1,1,3),(0,2,0),(-2,0,0) | (0,0,0) |
| 2.4 | P$\bar{1}$ | C5(1)P1(1) | (a,0,0,b,0,0,c) | (1,1,3),(0,2,0),(-2,0,0) | (0,1/2,0) |
| 8.34 | Cm' | C6(1)P1(1) | (a,-0.268a,b,-0.268b,0,0,c) | (-2,-2,0),(0,0,6),(-2,0,0) | (-1/12,-1/12,1/4) |
| 9.39 | Cc' | C7(1)P1(1) | (a,-0.268a,b,3.732b,0,0,c) | (-2,-2,0),(0,0,6),(-2,0,0) | (-7/12,-7/12,7/4) |
| 8.34 | Cm' | C8(1)P1(1) | (a,b,a,b,0,0,c) | (2,-2,0),(2,2,0),(-1,1,3) | (0,0,0) |
| 9.37 | Cc | C9(1)P1(1) | (a,b,b,-a,0,0,c) | (2,2,0),(-2,2,0),(1,1,3) | (0,3/2,0) |
| 156.51 | P3m'1 | C10(1)P1(1) | (a,b,a,b,a,b,c) | (0,-2,0),(2,2,0),(0,0,6) | (0,0,0) |
| 12.62 | C2'/m' | C11(1)P1(1) | (a,0,b,0,a,0,c) | (-2,-4,0),(2,0,0),(0,0,6) | (0,0,0) |
| 12.62 | C2'/m' | C12(1)P1(1) | (0,a,b,0,0,-a,c) | (-2,-4,0),(2,0,0),(0,0,6) | (1/2,0,0) |
| 38.191 | Am'm'2 | C13(1)P1(1) | (a,-0.268a,b,-0.268b,b,-0.268b,c) | (0,0,6),(0,2,0),(-4,-2,0) | (0,0,1/4) |
| 39.199 | Ab'm'2 | C14(1)P1(1) | (a,-0.268a,b,3.732b,-b,-3.732b,c) | (0,0,6),(0,2,0),(-4,-2,0) | (0,0,1/4) |
| 144.4 | P3$_1$ | C15(1)P1(1) | (a,b,-0.500a-0.866b,0.866a-0.500b,-0.500a+0.866b,-0.866a-0.500b,c) | (2,2,0),(-2,0,0),(0,0,6) | (0,0,0) |
| 5.15 | C2' | S1(1)P1(1) | (a,b,c,0,a,-b,d) | (-2,-4,0),(2,0,0),(0,0,6) | (0,0,0) |
| 5.13 | C2 | S2(1)P1(1) | (a,-0.268a,b,c,0.866b-0.500c,-0.500b-0.866c,d) | (0,-2,0),(4,2,0),(0,0,6) | (0,0,1/4) |
| 2.4 | P$\bar{1}$ | S3(1)P1(1) | (a,0,b,0,c,0,d) | (0,0,6),(0,2,0),(-2,0,0) | (0,0,0) |
| 2.4 | P$\bar{1}$ | S4(1)P1(1) | (a,0,0,b,0,c,d) | (0,0,6),(0,2,0),(-2,0,0) | (0,1/2,0) |
| 6.2 | Pm' | S5(1)P1(1) | (a,-0.268a,b,-0.268b,c,-0.268c,d) | (-2,0,0),(0,0,6),(0,2,0) | (0,0,1/4) |
| 7.26 | Pc' | S6(1)P1(1) | (a,-0.268a,b,3.732b,c,3.732c,d) | (-2,0,0),(0,0,6),(0,2,0) | (0,0,1/4) |
| 1.1 | P1 | 4D1(1)P1(1) | (a,b,c,d,0,0,e) | (1,1,3),(0,2,0),(-2,0,0) | (0,0,0) |
| 8.34 | Cm' | 4D2(1)P1(1) | (a,b,c,d,a,b,e) | (-2,-4,0),(2,0,0),(0,0,6) | (0,0,0) |
| 1.1 | P1 | 6D1(1)P1(1) | (a,b,c,d,e,f,g) | (0,0,6),(0,2,0),(-2,0,0) | (0,0,0) |

**S3.2. U1 and gM5+**

**Table S9**  Magnetic subgroups arising from coupling of irreps $m\Gamma_5^+$ and U1(1/2,0,1/6), with respect to the parent space group P6$_3$/mmc.

| Space Group Number | Space group | OPD Name | OPD Vector | Basis Vectors | Origin |
|---|---|---|---|---|---|
| 12.58 | C2/m | P1(1)P1(1) | (a,0,0,0,0,0,b,-1.732b) | (2,1,-6),(0,1,0),(2,1,0) | (0,0,0) |
| 44.229 | Imm2 | P2(1)P1(1) | (a,-0.268a,0,0,0,0,b,-1.732b) | (0,1,0),(0,0,6),(2,1,0) | (0,0,1/4) |
| 12.58 | C2/m | P3(1)P1(3) | (a,0,a,0,0,0,b,1.732b) | (2,-2,0),(2,2,0),(-1,1,3) | (0,0,0) |
| 15.85 | C2/c | P4(1)P1(3) | (a,0,0,a,0,0,b,1.732b) | (2,2,0),(-2,2,0),(1,1,3) | (0,1/2,0) |
| 43.224 | Fdd2 | P5(1)P1(3) | (a,-0.268a,0.268a,a,0,0,b,1.732b) | (-2,2,0),(0,0,6),(2,2,0) | (5/6,5/6,5/2) |
| 42.219 | Fmm2 | P6(1)P1(3) | (a,-0.268a,a,-0.268a,0,0,b,1.732b) | (2,2,0),(0,0,6),(2,-2,0) | (1/12,-1/12,1/4) |
| 8.32 | Cm | C1(1)P1(1) | (a,b,0,0,0,0,c,-1.732c) | (2,1,-6),(0,1,0),(2,1,0) | (0,0,0) |
| 5.13 | C2 | C2(1)P1(3) | (a,b,a,-b,0,0,c,1.732c) | (2,-2,0),(2,2,0),(-1,1,3) | (0,0,0) |
| 5.13 | C2 | C3(1)P1(3) | (a,b,0.866a-0.500b,-0.500a-0.866b,0,0,c,1.732c) | (2,2,0),(-2,2,0),(1,1,3) | (1/12,23/12,1/4) |
| 8.32 | Cm | C8(1)P1(3) | (a,b,a,b,0,0,c,1.732c) | (2,-2,0),(2,2,0),(-1,1,3) | (0,0,0) |
| 9.37 | Cc | C9(1)P1(3) | (a,b,b,-a,0,0,c,1.732c) | (2,2,0),(-2,2,0),(1,1,3) | (0,3/2,0) |
| 12.58 | C2/m | C11(1)P1(2) | (a,0,b,0,a,0,-2c,0) | (-2,-4,0),(2,0,0),(0,0,6) | (0,0,0) |
| 12.58 | C2/m | C12(1)P1(2) | (0,a,b,0,0,-a,-2c,0) | (-2,-4,0),(2,0,0),(0,0,6) | (1/2,0,0) |
| 38.187 | Amm2 | C13(1)P1(1) | (a,-0.268a,b,-0.268b,b,-0.268b,c,-1.732c) | (0,0,6),(0,2,0),(-4,-2,0) | (0,0,1/4) |
| 39.195 | Abm2 | C14(1)P1(1) | (a,-0.268a,b,3.732b,-b,-3.732b,c,-1.732c) | (0,0,6),(0,2,0),(-4,-2,0) | (0,0,1/4) |
| 5.13 | C2 | S1(1)P1(2) | (a,b,c,0,a,-b,-2d,0) | (-2,-4,0),(2,0,0),(0,0,6) | (0,0,0) |
| 5.13 | C2 | S2(1)P1(1) | (a,-0.268a,b,c,0.866b-0.500c,-0.500b-0.866c,d,-1.732d) | (0,-2,0),(4,2,0),(0,0,6) | (0,0,1/4) |
| 8.32 | Cm | 4D2(1)P1(2) | (a,b,c,d,a,b,-2e,0) | (-2,-4,0),(2,0,0),(0,0,6) | (0,0,0) |
| 12.62 | C2'/m' | P1(1)P2(1) | (a,0,0,0,0,0,b,0.577b) | (2,1,-6),(0,1,0),(2,1,0) | (0,0,0) |
| 44.231 | Im'm2' | P2(1)P2(1) | (a,-0.268a,0,0,0,0,b,0.577b) | (0,1,0),(0,0,6),(2,1,0) | (0,0,1/4) |
| 12.62 | C2'/m' | P3(1)P2(3) | (a,0,a,0,0,0,-b,0.577b) | (2,-2,0),(2,2,0),(-1,1,3) | (0,0,0) |
| 15.89 | C2'/c' | P4(1)P2(3) | (a,0,0,a,0,0,-b,0.577b) | (2,2,0),(-2,2,0),(1,1,3) | (0,1/2,0) |
| 43.226 | Fd'd2' | P5(1)P2(3) | (a,-0.268a,0.268a,a,0,0,-b,0.577b) | (-2,2,0),(0,0,6),(2,2,0) | (5/6,5/6,5/2) |
| 42.221 | Fm'm2' | P6(1)P2(3) | (a,-0.268a,a,-0.268a,0,0,-b,0.577b) | (2,2,0),(0,0,6),(2,-2,0) | (1/12,-1/12,1/4) |
| 8.34 | Cm' | C1(1)P2(1) | (a,b,0,0,0,0,c,0.577c) | (2,1,-6),(0,1,0),(2,1,0) | (0,0,0) |
| 5.15 | C2' | C2(1)P2(3) | (a,b,a,-b,0,0,-c,0.577c) | (2,-2,0),(2,2,0),(-1,1,3) | (0,0,0) |
| 5.15 | C2' | C3(1)P2(3) | (a,b,0.866a-0.500b,-0.500a-0.866b,0,0,-c,0.577c) | (2,2,0),(-2,2,0),(1,1,3) | (1/12,23/12,1/4) |
| 8.34 | Cm' | C8(1)P2(3) | (a,b,a,b,0,0,-c,0.577c) | (2,-2,0),(2,2,0),(-1,1,3) | (0,0,0) |
| 9.39 | Cc' | C9(1)P2(3) | (a,b,b,-a,0,0,-c,0.577c) | (2,2,0),(-2,2,0),(1,1,3) | (0,3/2,0) |
| 12.62 | C2'/m' | C11(1)P2(2) | (a,0,b,0,a,0,0,-1.155c) | (-2,-4,0),(2,0,0),(0,0,6) | (0,0,0) |
| 12.62 | C2'/m' | C12(1)P2(2) | (0,a,b,0,0,-a,0,-1.155c) | (-2,-4,0),(2,0,0),(0,0,6) | (1/2,0,0) |
| 38.19 | Amm'2' | C13(1)P2(1) | (a,-0.268a,b,-0.268b,b,-0.268b,c,0.577c) | (0,0,6),(0,2,0),(-4,-2,0) | (0,0,1/4) |
| 39.198 | Abm'2' | C14(1)P2(1) | (a,-0.268a,b,3.732b,-b,-3.732b,c,0.577c) | (0,0,6),(0,2,0),(-4,-2,0) | (0,0,1/4) |
| 5.15 | C2' | S1(1)P2(2) | (a,b,c,0,a,-b,0,-1.155d) | (-2,-4,0),(2,0,0),(0,0,6) | (0,0,0) |
| 5.15 | C2' | S2(1)P2(1) | (a,-0.268a,b,c,0.866b-0.500c,-0.500b-0.866c,d,0.577d) | (0,-2,0),(4,2,0),(0,0,6) | (0,0,1/4) |

| SGN | Space group | OPD Name | OPD vector | Basis Vectors | Origin |
|---|---|---|---|---|---|
| 8.34 | Cm' | 4D2(1)P2(2) | (a,b,c,d,a,b,0,-1.155e) | (-2,-4,0),(2,0,0),(0,0,6) | (0,0,0) |
| 2.4 | P$\bar{1}$ | P1(1)C1(1) | (a,0,0,0,0,0,b,c) | (1,0,3),(0,1,0),(-2,0,0) | (0,0,0) |
| 8.32 | Cm | P2(1)C1(1) | (a,-0.268a,0,0,0,0,b,c) | (-2,0,0),(0,0,6),(0,1,0) | (-1/12,0,1/4) |
| 1.1 | P1 | C1(1)C1(1) | (a,b,0,0,0,0,c,d) | (1,0,3),(0,1,0),(-2,0,0) | (0,0,0) |
| 2.4 | P$\bar{1}$ | C4(1)C1(1) | (a,0,b,0,0,0,c,d) | (1,1,3),(0,2,0),(-2,0,0) | (0,0,0) |
| 2.4 | P$\bar{1}$ | C5(1)C1(1) | (a,0,0,b,0,0,c,d) | (1,1,3),(0,2,0),(-2,0,0) | (0,1/2,0) |
| 8.32 | Cm | C6(1)C1(1) | (a,-0.268a,b,-0.268b,0,0,c,d) | (-2,-2,0),(0,0,6),(-2,0,0) | (-1/12,-1/12,1/4) |
| 9.37 | Cc | C7(1)C1(1) | (a,-0.268a,b,3.732b,0,0,c,d) | (-2,-2,0),(0,0,6),(-2,0,0) | (-7/12,-7/12,7/4) |
| 2.4 | P$\bar{1}$ | S3(1)C1(1) | (a,0,b,0,c,0,d,e) | (0,0,6),(0,2,0),(-2,0,0) | (0,0,0) |
| 2.4 | P$\bar{1}$ | S4(1)C1(1) | (a,0,0,b,0,c,d,e) | (0,0,6),(0,2,0),(-2,0,0) | (0,1/2,0) |
| 6.18 | Pm | S5(1)C1(1) | (a,-0.268a,b,-0.268b,c,-0.268c,d,e) | (-2,0,0),(0,0,6),(0,2,0) | (0,0,1/4) |
| 7.24 | Pc | S6(1)C1(1) | (a,-0.268a,b,3.732b,c,3.732c,d,e) | (-2,0,0),(0,0,6),(0,2,0) | (0,0,1/4) |
| 1.1 | P1 | 4D1(1)C1(1) | (a,b,c,d,0,0,e,f) | (1,1,3),(0,2,0),(-2,0,0) | (0,0,0) |
| 1.1 | P1 | 6D1(1)C1(1) | (a,b,c,d,e,f,g,h) | (0,0,6),(0,2,0),(-2,0,0) | (0,0,0) |

## S3.3. 3C

### S3.3.1. U1

**Table S10** Order parameter components and unit cell configurations for subgroups of *P*6$_3$/*mmc* which can arise from phase transitions in which irrep U1(1/2,0,1/3) is the active representation.

| SGN | Space group | OPD Name | OPD vector | Basis Vectors | Origin |
|---|---|---|---|---|---|
| 58 | Pnnm | P1 | (a,0,0,0,0,0) | (0,0,3),(2,1,0),(0,1,0) | (0,0,0) |
| 59 | Pmmn | P2 | (0,a,0,0,0,0) | (0,0,3),(0,-1,0),(2,1,0) | (1/2,1/2,0) |

| | | | | | | |
|---|---|---|---|---|---|---|
| 164 | P$\bar{3}$m1 | P3 | (a,0,a,0,a,0) | (2,0,0),(0,2,0),(0,0,3) | (0,0,0) |
| 187 | P$\bar{6}$m2 | P4 | (a,-0.577a,a,-0.577a,a,-0.577a) | (0,-2,0),(2,2,0),(0,0,3) | (0,0,1/4) |
| 64 | Cmca | P5 | (a,0,0,0,a,0) | (2,0,0),(2,4,0),(0,0,3) | (0,0,0) |
| 63 | Cmcm | P6 | (0,a,0,0,0,-a) | (2,0,0),(2,4,0),(0,0,3) | (1/2,0,0) |
| 152 | P3$_1$21 | P7 | (a,1.732a,-2a,0,a,-1.732a) | (2,0,0),(0,2,0),(0,0,3) | (0,0,1/2) |
| 151 | P3$_1$12 | P8 | (a,0.577a,a,0.577a,0,-1.155a) | (0,-2,0),(2,2,0),(0,0,3) | (0,0,3/4) |
| 31 | Pmn2$_1$ | C1 | (a,b,0,0,0,0) | (0,-1,0),(2,1,0),(0,0,3) | (1/2,1/4,0) |
| 156 | P3m1 | C2 | (a,b,a,b,a,b) | (2,0,0),(0,2,0),(0,0,3) | (0,0,0) |
| 20 | C222$_1$ | C3 | (a,b,0,0,a,-b) | (2,0,0),(2,4,0),(0,0,3) | (1/2,0,0) |
| 14 | P2$_1$/c | C4 | (0,0,a,0,b,0) | (-2,0,0),(0,0,3),(0,2,0) | (0,0,0) |
| 11 | P2$_1$/m | C5 | (0,0,0,a,0,b) | (-2,0,0),(0,0,3),(0,2,0) | (0,1/2,0) |
| 14 | P2$_1$/c | C6 | (a,0,0,0,0,b) | (-2,0,0),(0,0,3),(2,2,0) | (0,1/2,0) |
| 12 | C2/m | C7 | (a,0,b,0,a,0) | (-2,-4,0),(2,0,0),(0,0,3) | (0,0,0) |
| 12 | C2/m | C8 | (0,a,b,0,0,-a) | (-2,-4,0),(2,0,0),(0,0,3) | (1/2,0,0) |
| 38 | Amm2 | C9 | (a,-0.577a,b,-0.577b,b,-0.577b) | (0,0,3),(0,2,0),(-4,-2,0) | (0,0,1/4) |
| 39 | Abm2 | C10 | (a,-0.577a,b,1.732b,-b,1.732b) | (0,0,3),(0,2,0),(-4,-2,0) | (0,0,1/4) |
| 36 | Cmc2$_1$ | C11 | (a,b,0,0,a,b) | (2,0,0),(2,4,0),(0,0,3) | (1/2,-1,0) |
| 144 | P3$_1$ | C12 | (a,b,-0.500a0.866b,0.866a-0.500b,-0.500a+0.866b,-0.866a-0.500b) | (2,2,0),(-2,0,0),(0,0,3) | (0,0,0) |
| 5 | C2 | S1 | (a,b,c,0,a,-b) | (-2,-4,0),(2,0,0),(0,0,3) | (0,0,0) |
| 5 | C2 | S2 | (a,-0.577a,b,c,0.500b0.866c,-0.866b0.500c) | (0,-2,0),(4,2,0),(0,0,3) | (0,0,1/4) |
| 2 | P$\bar{1}$ | S3 | (a,0,b,0,c,0) | (0,0,3),(0,2,0),(-2,0,0) | (0,0,0) |
| 2 | P$\bar{1}$ | S4 | (a,0,0,b,0,c) | (0,0,3),(0,2,0),(-2,0,0) | (0,1/2,0) |
| 6 | Pm | S5 | (a,-0.577a,b,-0.577b,c,-0.577c) | (-2,0,0),(0,0,3),(0,2,0) | (0,0,1/4) |
| 7 | Pc | S6 | (a,-0.577a,b,1.732b,c,1.732c) | (-2,0,0),(0,0,3),(0,2,0) | (0,0,1/4) |
| 4 | P2$_1$ | 4D1 | (0,0,a,b,c,d) | (-2,0,0),(0,0,3),(0,2,0) | (0,1/2,0) |
| 8 | Cm | 4D2 | (a,b,c,d,a,b) | (-2,-4,0),(2,0,0),(0,0,3) | (0,0,0) |
| 1 | P1 | 6D1 | (a,b,c,d,e,f) | (0,0,3),(0,2,0),(-2,0,0) | (0,0,0) |

### S3.3.2. U1 and mGm4+

**Table S11** Magnetic subgroups arising from coupling of irreps m$\Gamma_4^+$ and U1(1/2,0,1/3), with respect to the parent space group P6$_3$/mmc.

| SGN.MSGN | Subgroup | OPD | OPD vector | Basis Vectors | Origin |
|---|---|---|---|---|---|
| 58.398 | Pnn'm' | P1(1)P1(1) | (a,0,0,0,0,b) | (-2,-1,0),(0,0,3),(0,1,0) | (0,0,0) |
| 59.409 | Pm'm'n | P2(1)P1(1) | (0,a,0,0,0,b) | (0,0,3),(0,-1,0),(2,1,0) | (1/2,1/2,0) |
| 164.89 | P$\bar{3}$m'1 | P3(1)P1(1) | (a,0,a,0,a,0,b) | (0,-2,0),(2,2,0),(0,0,3) | (0,0,0) |
| 187.211 | P$\bar{6}$'m'2 | P4(1)P1(1) | (a,-0.577a,a,-0.577a,a,-0.577a,b) | (2,2,0),(-2,0,0),(0,0,3) | (0,0,1/4) |
| 64.476 | Cm'ca' | P5(1)P1(1) | (a,0,0,0,a,0,b) | (2,0,0),(2,4,0),(0,0,3) | (0,0,0) |
| 63.464 | Cm'cm' | P6(1)P1(1) | (0,a,0,0,0,-a,b) | (2,0,0),(2,4,0),(0,0,3) | (1/2,0,0) |
| 152.35 | P3$_1$2'1 | P7(1)P1(1) | (a,1.732a,-2a,0,a,-1.732a,b) | (0,-2,0),(2,2,0),(0,0,3) | (0,0,0) |
| 151.29 | P3$_1$12 | P8(1)P1(1) | (a,0.577a,-a,0.577a,0,-1.155a,b) | (2,2,0),(-2,0,0),(0,0,3) | (0,0,1/4) |
| 31.125 | Pm'n2$_1$' | C1(1)P1(1) | (a,b,0,0,0,0,c) | (0,-1,0),(2,1,0),(0,0,3) | (1/2,1/4,0) |
| 156.51 | P3m'1 | C2(1)P1(1) | (a,b,a,b,a,b,c) | (0,-2,0),(2,2,0),(0,0,3) | (0,0,0) |
| 20.34 | C22'2$_1$' | C3(1)P1(1) | (a,b,0,0,a,-b,c) | (2,4,0),(-2,0,0),(0,0,3) | (3/2,2,3/4) |
| 14.79 | P2$_1$'/c' | C4(1)P1(1) | (0,0,a,0,b,0,c) | (-2,0,0),(0,0,3),(0,2,0) | (0,0,0) |
| 11.54 | P2$_1$'/m' | C5(1)P1(1) | (0,0,0,a,0,b,c) | (-2,0,0),(0,0,3),(0,2,0) | (0,1/2,0) |
| 14.79 | P2$_1$'/c' | C6(1)P1(1) | (a,0,0,0,0,b,c) | (-2,0,0),(0,0,3),(2,2,0) | (0,1/2,0) |
| 12.62 | C2'/m' | C7(1)P1(1) | (a,0,b,0,a,0,c) | (-2,-4,0),(2,0,0),(0,0,3) | (0,0,0) |
| 12.62 | C2'/m' | C8(1)P1(1) | (0,a,b,0,0,-a,c) | (-2,-4,0),(2,0,0),(0,0,3) | (1/2,0,0) |
| 38.191 | Am'm'2 | C9(1)P1(1) | (a,-0.577a,b,-0.577b,b,-0.577b,c) | (0,0,3),(0,2,0),(-4,-2,0) | (0,0,1/4) |
| 39.199 | Ab'm'2 | C10(1)P1(1) | (a,-0.577a,b,1.732b,-b,-1.732b,c) | (0,0,3),(0,2,0),(-4,-2,0) | (0,0,1/4) |

| 36.174 | Cm'c2$_1$' | C11(1)P1(1) | (a,b,0,0,a,b,c) | (2,0,0),(2,4,0),(0,0,3) | (1/2,-1,0) |
| 144.4 | P3$_1$ | C12(1)P1(1) | (a,b,-0.500a-0.866b,0.866a-0.500b,-0.500a+0.866b,-0.866a-0.500b,c) | (2,2,0),(-2,0,0),(0,0,3) | (0,0,0) |
| 5.15 | C2' | S1(1)P1(1) | (a,b,c,0,a,-b,d) | (-2,-4,0),(2,0,0),(0,0,3) | (0,0,0) |
| 5.13 | C2 | S2(1)P1(1) | (a,-0.577a,b,c,0.500b-0.866c,-0.866b-0.500c,d) | (0,-2,0),(4,2,0),(0,0,3) | (0,0,1/4) |
| 2.4 | P$\bar{1}$ | S3(1)P1(1) | (a,0,b,0,c,0,d) | (0,0,3),(0,2,0),(-2,0,0) | (0,0,0) |
| 2.4 | P$\bar{1}$ | S4(1)P1(1) | (a,0,0,b,0,c,d) | (0,0,3),(0,2,0),(-2,0,0) | (0,1/2,0) |
| 6.2 | Pm' | S5(1)P1(1) | (a,-0.577a,b,-0.577b,c,-0.577c,d) | (-2,0,0),(0,0,3),(0,2,0) | (0,0,1/4) |
| 7.26 | Pc' | S6(1)P1(1) | (a,-0.577a,b,1.732b,c,1.732c,d) | (-2,0,0),(0,0,3),(0,2,0) | (0,0,1/4) |
| 4.9 | P2$_1$' | 4D1(1)P1(1) | (0,0,a,b,c,d,e) | (-2,0,0),(0,0,3),(0,2,0) | (0,1/2,0) |
| 8.34 | Cm' | 4D2(1)P1(1) | (a,b,c,d,a,b,e) | (-2,-4,0),(2,0,0),(0,0,3) | (0,0,0) |
| 1.1 | P1 | 6D1(1)P1(1) | (a,b,c,d,e,f,g) | (0,0,3),(0,2,0),(-2,0,0) | (0,0,0) |

**S3.3.3. U1 and mGm5+**

**Table S12** Magnetic subgroups arising from coupling of irreps m$\Gamma_5^+$ and U1(1/2,0,1/3), with respect to the parent space group P6$_3$/mmc.

| SGN.MSGN | Subgroup | OPD | OPD vector | Basis Vectors | Origin |
|---|---|---|---|---|---|
| 58.393 | Pnnm | P1(1)P1(1) | (a,0,0,0,0,b,-1.732b) | (-2,-1,0),(0,0,3),(0,1,0) | (0,0,0) |
| 59.405 | Pmmn | P2(1)P1(1) | (0,a,0,0,0,b,-1.732b) | (0,1,0),(0,0,3),(2,1,0) | (1/2,0,0) |
| 64.469 | Cmca | P5(1)P1(2) | (a,0,0,0,a,0,-2b,0) | (2,0,0),(2,4,0),(0,0,3) | (0,0,0) |
| 63.457 | Cmcm | P6(1)P1(2) | (0,a,0,0,0,-a,-2b,0) | (2,0,0),(2,4,0),(0,0,3) | (1/2,0,0) |
| 31.123 | Pmn2$_1$ | C1(1)P1(1) | (a,b,0,0,0,0,c,-1.732c) | (0,-1,0),(2,1,0),(0,0,3) | (1/2,1/4,0) |
| 20.31 | C222$_1$ | C3(1)P1(2) | (a,b,0,0,a,-b,-2c,0) | (2,0,0),(2,4,0),(0,0,3) | (1/2,0,0) |
| 12.58 | C2/m | C7(1)P1(2) | (a,0,b,0,a,0,-2c,0) | (-2,-4,0),(2,0,0),(0,0,3) | (0,0,0) |
| 12.58 | C2/m | C8(1)P1(2) | (0,a,b,0,0,-a,-2c,0) | (-2,-4,0),(2,0,0),(0,0,3) | (1/2,0,0) |
| 38.187 | Amm2 | C9(1)P1(1) | (a,-0.577a,b,-0.577b,b,-0.577b,c,-1.732c) | (0,0,3),(0,2,0),(-4,-2,0) | (0,0,1/4) |
| 39.195 | Abm2 | C10(1)P1(1) | (a,-0.577a,b,1.732b,-b,-1.732c) | (0,0,3),(0,2,0),(-4,-2,0) | (0,0,1/4) |
| 36.172 | Cmc2$_1$ | C11(1)P1(2) | (a,b,0,0,a,b,-2c,0) | (2,0,0),(2,4,0),(0,0,3) | (1/2,-1,0) |
| 5.13 | C2 | S1(1)P1(2) | (a,b,c,0,a,-b,-2d,0) | (-2,-4,0),(2,0,0),(0,0,3) | (0,0,0) |
| 5.13 | C2 | S2(1)P1(1) | (a,-0.577a,b,c,0.500b-0.866c,-0.866b-0.500c,d,-1.732d) | (0,-2,0),(4,2,0),(0,0,3) | (0,0,1/4) |
| 8.32 | Cm | 4D2(1)P1(2) | (a,b,c,d,a,b,-2e,0) | (-2,-4,0),(2,0,0),(0,0,3) | (0,0,0) |
| 58.398 | Pnn'm' | P1(1)P2(1) | (a,0,0,0,0,0,b,0.577b) | (0,0,3),(2,1,0),(0,1,0) | (0,0,0) |
| 59.41 | Pmm'n' | P2(1)P2(1) | (0,a,0,0,0,0,b,0.577b) | (0,0,3),(0,-1,0),(2,1,0) | (1/2,1/2,0) |
| 64.474 | Cm'c'a | P5(1)P2(2) | (a,0,0,0,a,0,0,-1.155b) | (2,0,0),(2,4,0),(0,0,3) | (0,0,0) |
| 63.462 | Cm'c'm | P6(1)P2(2) | (0,a,0,0,0,-a,0,-1.155b) | (2,0,0),(2,4,0),(0,0,3) | (1/2,0,0) |
| 31.127 | Pm'n'2$_1$ | C1(1)P2(1) | (a,b,0,0,0,0,c,0.577c) | (0,-1,0),(2,1,0),(0,0,3) | (1/2,1/4,0) |
| 20.33 | C2'2'2$_1$ | C3(1)P2(2) | (a,b,0,0,a,-b,0,-1.155c) | (2,0,0),(2,4,0),(0,0,3) | (1/2,0,0) |
| 12.62 | C2'/m' | C7(1)P2(2) | (a,0,b,0,a,0,0,-1.155c) | (-2,-4,0),(2,0,0),(0,0,3) | (0,0,0) |
| 12.62 | C2'/m' | C8(1)P2(2) | (0,a,b,0,0,-a,0,-1.155c) | (-2,-4,0),(2,0,0),(0,0,3) | (1/2,0,0) |
| 38.19 | Amm'2' | C9(1)P2(1) | (a,-0.577a,b,-0.577b,b,-0.577b,c,0.577c) | (0,0,3),(0,2,0),(-4,-2,0) | (0,0,1/4) |
| 39.198 | Abm'2' | C10(1)P2(1) | (a,-0.577a,b,1.732b,-b,-1.732b,c,0.577c) | (0,0,3),(0,2,0),(-4,-2,0) | (0,0,1/4) |
| 36.176 | Cm'c'2$_1$ | C11(1)P2(2) | (a,b,0,0,a,b,0,-1.155c) | (2,0,0),(2,4,0),(0,0,3) | (1/2,-1,0) |
| 5.15 | C2' | S1(1)P2(2) | (a,b,c,0,a,-b,0,-1.155d) | (-2,-4,0),(2,0,0),(0,0,3) | (0,0,0) |
| 5.15 | C2' | S2(1)P2(1) | (a,-0.577a,b,c,0.500b-0.866c,-0.866b-0.500c,d,0.577d) | (0,-2,0),(4,2,0),(0,0,3) | (0,0,1/4) |
| 8.34 | Cm' | 4D2(1)P2(2) | (a,b,c,d,a,b,0,-1.155e) | (-2,-4,0),(2,0,0),(0,0,3) | (0,0,0) |
| 14.75 | P2$_1$/c | P1(1)C1(1) | (a,0,0,0,0,0,b,c) | (0,1,0),(0,0,3),(2,0,0) | (0,0,0) |
| 11.5 | P2$_1$/m | P2(1)C1(1) | (0,a,0,0,0,0,b,c) | (-2,0,0),(0,0,3),(0,1,0) | (-1/2,0,0) |
| 4.7 | P2$_1$ | C1(1)C1(1) | (a,b,0,0,0,0,c,d) | (-2,0,0),(0,0,3),(0,1,0) | (-1/2,0,0) |
| 14.75 | P2$_1$/c | C4(1)C1(1) | (0,0,a,0,b,0,c,d) | (-2,0,0),(0,0,3),(0,2,0) | (0,0,0) |
| 11.5 | P2$_1$/m | C5(1)C1(1) | (0,0,0,a,0,b,c,d) | (-2,0,0),(0,0,3),(0,2,0) | (0,1/2,0) |
| 14.75 | P2$_1$/c | C6(1)C1(1) | (a,0,0,0,0,b,c,d) | (-2,0,0),(0,0,3),(2,2,0) | (0,1/2,0) |
| 2.4 | P$\bar{1}$ | S3(1)C1(1) | (a,0,b,0,c,0,d,e) | (0,0,3),(0,2,0),(-2,0,0) | (0,0,0) |
| 2.4 | P$\bar{1}$ | S4(1)C1(1) | (a,0,0,b,0,c,d,e) | (0,0,3),(0,2,0),(-2,0,0) | (0,1/2,0) |
| 6.18 | Pm | S5(1)C1(1) | (a,-0.577a,b,-0.577b,c,-0.577c,d,e) | (-2,0,0),(0,0,3),(0,2,0) | (0,0,1/4) |
| 7.24 | Pc | S6(1)C1(1) | (a,-0.577a,b,1.732b,c,1.732c,d,e) | (-2,0,0),(0,0,3),(0,2,0) | (0,0,1/4) |
| 4.7 | P2$_1$ | 4D1(1)C1(1) | (0,0,a,b,c,d,e,f) | (-2,0,0),(0,0,3),(0,2,0) | (0,1/2,0) |
| 1.1 | P1 | 6D1(1)C1(1) | (a,b,c,d,e,f,g,h) | (0,0,3),(0,2,0),(-2,0,0) | (0,0,0) |

**Acknowledgements** The authors acknowledge funding from the Leverhulme Foundation, grant number RPG-2016-298.